\documentclass[
	aps, prb,
	superscriptaddress, 
	showkeywords,
	reprint,
	longbibliography,
	amsmath,amssymb,floatfix.xcolor=table,
	]{revtex4-2}

\usepackage{graphicx}
\usepackage{siunitx}
\usepackage{textcomp}

\usepackage{booktabs}
\usepackage{multirow}
\usepackage{longtable}
\usepackage{tabularx}
\usepackage{array}

\renewcommand*{\epsilon}{\varepsilon}
\renewcommand*{\vec}[1]{\mathbf{#1}}

\DeclareTextCommand{\permil}{T1}{\%\char24}

\newcommand{\half}{\frac{1}{2}}
\newcommand{\sqrthalf}{\frac{1}{\sqrt{2}}}
	
\usepackage{xcolor}

\newcommand{\correspondingemail}{Now at Loughborough University, UK, n.leo@lboro.ac.uk}

\newcommand{\csic}{Instituto de Nanociencia y Materiales de Arag\'on (INMA), CSIC-Universidad de Zaragoza, 50009 Zaragoza, Spain}
\newcommand{\nanogune}{CIC nanoGUNE BRTA, 20018 Donostia - San Sebasti\'{a}n, Spain}

\newcommand{\riec}{Research Institute of Electrical Communication (RIEC), Tohoku University, Sendai 980-8577, Japan.}
\newcommand{\gsis}{Graduate School of Information Sciences (GSIS), Tohoku University, Sendai 980-8579, Japan.}
\newcommand{\csis}{Center for Science and Innovation in Spintronics (CSIS), Tohoku University, Sendai 980-8577, Japan.}

\newcommand{\sigmai}{Sigma-i Co. Ltd., Minato-ku, Tokyo 108-0075, Japan.}

\newcommand{\ciies}{Center for Innovative Integrated Electronic Systems (CIES), Tohoku University, Sendai 980-0845, Japan.}
\newcommand{\wpi}{WPI Advanced Institute for Materials Research, Tohoku University, Sendai 980-8577, Japan.}
\newcommand{\inamori}{Inamori Research Insitute for Science, Kyoto 600-8411, Japan.}

\begin{document}

\title{Magnetic order in nanoscale gyroid networks}

\author{Ami S.\ Koshikawa}
\affiliation{\gsis}

\author{Justin Llandro}
\affiliation{\riec}
\affiliation{\csis}

\author{Masayuki Ohzeki}
\affiliation{\gsis}
\affiliation{International Research Frontiers Initiative, Tokyo Institute of Technology, Shibaura, Minato-ku, Tokyo 108-0023, Japan}
\affiliation{\sigmai}

\author{Shunsuke Fukami}
\affiliation{\riec}
\affiliation{\csis}
\affiliation{\ciies}
\affiliation{\wpi}
\affiliation{\inamori}

\author{Hideo Ohno}
\affiliation{\riec}
\affiliation{\csis}
\affiliation{\ciies}
\affiliation{\wpi}

\author{Na\"emi Leo}
\email{\correspondingemail}
\affiliation{\csic}
\affiliation{\nanogune} 

\begin{abstract}

Three-dimensional magnetic metamaterials feature interesting phenomena that arise from a delicate interplay of material properties, local anisotropy, curvature, and connectivity.
%
A particularly interesting magnetic lattice that combines these aspects is that of nanoscale gyroids, with a highly-interconnected chiral network with local three-connectivity reminiscent of three-dimensional artificial spin ices. 
%
Here, we use finite-element micromagnetic simulations to elucidate the anisotropic behaviour of nanoscale nickel gyroid networks at applied fields and at remanence. 
%
We simplify the description of the micromagnetic spin states with a macrospin model to explain the anistropic global response, to quantify the extent of ice-like correlations, and to discuss qualitative features of the anisotropic magnetoresistance in the three-dimensional network.
%
Our results demonstrate the large variability of the magnetic order in extended gyroid networks, 
%
which might enable future spintronic functionalities, including neuromorphic computing and non-reciprocal transport.

\end{abstract}

\keywords{3D nanomagnetism, artificial spin ice, nanowire networks, magnetotransport, geometrical effects}

\maketitle

Networks of interacting nanomagnetic wires offer insight into emergent phenomena and functionalities arising from the underlying geometrical design and local connectivity.
A well-studied class of these networks are two-dimensional artificial spin ices (ASI) and magnonic crystals \cite{2019Skjaervoe, 2020Gliga, 2021Barman}, which allowed the observation via imaging or magnetotransport of ice-like low-energy states \cite{2012Branford, 2016Canals, 2017Gartside} and  monopole-like excitations \cite{2010Mengotti, 2013Rougemaille, 2017Le}.
Because of the stochastic behaviour and large reconfigurability of interacting interconnected lattices, such magnetic metamaterials have also been proposed to be used for neuromorphic-inspired unconventional computational tasks \cite{2021Dawidek, 2022Gartside, 2022Bhattacharya}.

Extending the study of emergent magnetic phenomena from planar two-dimensional to three-dimensional lattices promises novel functionalities \cite{2014Chern_a, 2016Perrin, 2017Fernandez, 2022Makarov_a}, related to magnetochiral effects in curvilinear geometries \cite{2016Streubel, 2018Volkov, 2020Sheka}, fast magnetisation dynamics \cite{2021Sahoo, 2022Cheenikundil, 2022Koerber, 2022Skoric}, and network topologies with dense connections to distant neighbours \cite{2021May, 2021Koraltan, 2022Pip}.
Notable examples of magnetic three-dimensional networks studied so far include inverse opals \cite{2011Grigoryeva, 2016Shishkin, 2019Mistonov, 2023Rana}, magnetic buckyballs \cite{2021Cheenikundil, 2022Cheenikundil}, and single-diamond lattices \cite{2021May, 2021Sahoo}. In these studies, the connecting struts are usually several \SI{100}{nm} long and thus much larger than typical magnetic length scales, weakening possible curvilinear magnetic effects expected in truly nanoscale 3D networks.

Gyroid structures grown by polymer self-assembly feature a highly interconnected three-dimensional network, a global chiral structure and a lattice periodicity in the order of a few tens of nanometres. Recent studies on photonic gyroids demonstrated selective reflection of circularly polarised light \cite{2013Turner} and the emergence of Weyl points \cite{2005Koshino,2013Lu, 2015Lu}. 
With respect to magnetism, the local curvature of the gyroid is large enough to support a sizeable geometrical Dyzaloshinskii-Moriya interaction \cite{2016Streubel, 2018Volkov} and its inherent chirality can give rise to emergent non-reciprocal effects \cite{2016Seki, 2018Tokura, 2020Barman}, such as electrical magneto-chiral anisotropy \cite{2001Rikken, 2021Atzori}.
In our previous work \cite{2020Llandro}, we imaged magnetic states of nanoscale gyroids using electron holography and observed complex magnetic states. However, the local spin anisotropy has not been elucidated and therefore possible ice-like correlations in magnetic gyroids have not yet been quantified so far.

Here, we use finite-element micromagnetic simulations to show that the field-driven and relaxed spin configurations of nanoscale nickel gyroids feature complex magnetic states arising from the non-trivial local anisotropy and the three-connectivity of the gyroid lattice, including the emergence of spin chiral effects. 
We discuss how the description of local spin order can be simplified using a macrospin model. We furthermore illustrate how the anisotropic magnetoresistance in finite-size gyroid networks, which, due to the inherently non-coplanar spin configuration as well as the 3D network connectivity, shows distinct behaviour compared to the response of bulk or planar devices.
%
Our results underline the complexity of magnetic order in nanoscale 3D gyroids with inherently non-coplanar and frustrated spin order. These properties make gyroid networks ideal candidates for future studies of non-reciprocal effects or as platform for probabilistic and neuromorphic computing schemes.

\begin{figure*}[t]
	\centering
	\includegraphics[width=178mm]{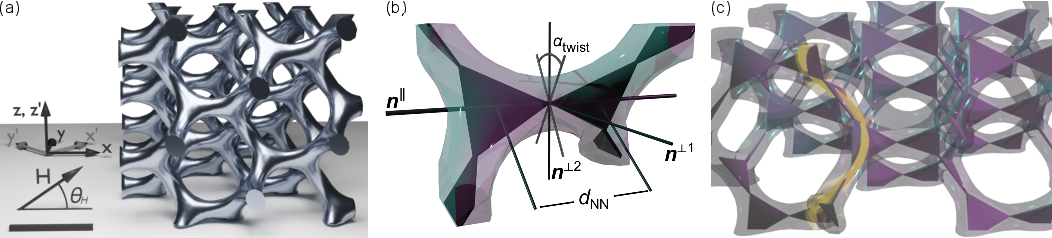}
	\caption[Gyroid geometry]{ %
		\textbf{Gyroid geometry.}
		(a)~Gyroid structure used for micromagnetic simulations, with directions of cubic coordinate system ($x^\prime, y^\prime, z^\prime$) and simulation coordinates ($x, y, z$). A magnetic field was applied in the $xz$ plane at angle $\theta_M$ to the $x$ direction. Scale bar measures \SI{50}{nm}.
		(b)~Definition of local coordinate axes $\hat{\mathbf{n}}_i^\parallel$, $\hat{\mathbf{n}}_i^{\perp 1}$, $\hat{\mathbf{n}}_i^{\perp 2}$, with $\vec{n}_i^\parallel$ of struts connecting neighbouring vertices with distance $d_\text{NN}$. The planes of neighbouring triangular plaquettes, with nodes centred on each strut, have a relative twist of $\alpha_\text{twist}\approx\ang{70.5}$.
		(c)~The gyroid can be represented by a highly-connected network of corner-sharing triangular plaquettes. The yellow line highlights the tightest helix path used to calculate the maximum spin canting due to the geometric Dzyaloshinskii-Moriya interaction. 
	}
	\label{fig:gyroid_geometry}
\end{figure*}

\section{The gyroid geometry}

A single gyroid shown in Fig.~\ref{fig:gyroid_geometry}(a), derived from the Schoen \textit{G} triply-periodic minimal surface \cite{1970Schoen}, is a 3D periodic network of connected struts which form chiral triple junctions.
While gyroid photonic crystals are found in nature in the wings of some butterflies \cite{2007Michielsen,2010Saranathan}, nanoscale gyroids with lattice periodicities $a$ in the range of \SI{40}{nm} to \SI{100}{nm} can be grown with large structural coherence over a few hundred micrometres by self-assembly of di- and tri-block co-polymer templates \cite{2012Vignolini, 2015Dolan, 2015Wu_a}.
Selective etching followed by metal electrodeposition into the remaining scaffold results in single-gyroid network nanostructures with volume fill fractions $f_V$ between 10\% and 30\% \cite{2018Dolan}.

Many of the interesting physical phenomena in gyroids are related to its inherent chirality, described by the cubic-centered space group \textit{I}4$_1$32 (which allows for uniquely left- or right-handed gyroid structures), and the connectivity, mathematically also described as the \textit{srs} net or \textit{K}$_4$ crystal \cite{2008Hyde_english, 2019Mizuno}. 
Vertices are connected to their neighbours by struts of length $d_\text{NN}=a/\sqrt{8}$, with $a$ being the cubic lattice constant. Each cubic unit cell contains eight individual vertices and 18 struts.
For each of the six strut directions, a local coordinate system can be defined as shown in Fig.~\ref{fig:gyroid_geometry}(b) and summarised in Tab.~\ref{tab:strut-directions}, with $\hat{\mathbf{n}}_i^\parallel$ denoting the main strut direction and neighbouring triangle planes rotated by $\alpha_\text{twist}\approx\ang{70.5}$.

Due to its underlying three-connectivity, the gyroid network can also be represented by corner-sharing triangles, as shown in Fig.~\ref{fig:gyroid_geometry}(c), reminiscent of geometrically-frustrated magnetic systems that promote complex spin states, zero-temperature entropy and other interesting emergent properties like magnetic monopoles \cite{1950Wannier, 2004Diep, Lacroix11}. 
Furthermore, a multitude of possible paths through the network exist. These include gyrating channels, such as the one highlighted yellow in Fig.~\ref{fig:gyroid_geometry}(c) corresponding to the tightest possible helix path through the gyroid with radius $r_H=a/(4\sqrt{2})$ and periodicity $p_H=a$.

\begin{table}[b!]
	\centering
	\caption{%
		\textbf{Local coordinate systems} for the six unique strut directions in the gyroid network, expressed with respect to coordinates $x, y, z$ of the simulation coordinate systems. 
		The local normalised direction vectors $\hat{\mathbf{n}}_i^\parallel$, $\hat{\mathbf{n}}_i^{\perp 1}$ and $\hat{\mathbf{n}}_i^{\perp 2}=\vec{n}_i^\parallel\times\vec{n}_i^{\perp 1}$ form a right-handed system, in agreement with the overall right-handed chirality of our gyroid lattice.
	}
	\label{tab:strut-directions}
	\vspace{1ex}
	\begin{tabular}{c||c|c|c}
		strut $i$ & $\hat{\mathbf{n}}^\parallel_i$ & $\hat{\mathbf{n}}^{\perp1}_i$ & $\hat{\mathbf{n}}^{\perp2}_i$ \\ \hline \hline
		1 & $(+1, 0, 0)$ & $(0, 0, -1)$ & $(0, +1, 0)$\\
		2 & $(0, +1, 0)$ & $(0, 0, -1)$ & $(+1, 0, 0)$\\
		3 & $(+\half, +\half, +\sqrthalf)$ & $(+\sqrthalf, -\sqrthalf, 0)$ & $(+\half, +\half, -\sqrthalf)$ \\
		4 & $(-\half, -\half, +\sqrthalf)$ & $(+\sqrthalf, -\sqrthalf, 0)$ & $(+\half, +\half, +\sqrthalf)$ \\
		5 & $(+\half, -\half, +\sqrthalf)$ & $(-\sqrthalf, -\sqrthalf, 0)$ & $(+\half, -\half, -\sqrthalf)$ \\
		6 & $(-\half, +\half, +\sqrthalf)$ & $(+\sqrthalf, +\sqrthalf, 0)$ & $(+\half, -\half, +\sqrthalf)$ \\
	\end{tabular}
\end{table}

\section{Micromagnetic Simulations}

Inspired by single-gyroids networks grown by self-assembly and studied by electron holography \cite{2020Llandro}, in this work we focus on gyroids with a cubic lattice constant $a=\SI{65}{nm}$ and a volume fraction of $f_V=17\%$. We use a coordinate system ($x, y, z$) rotated by \ang{45} around the direction $z^\prime$ (with cubic crystallographic coordinates $x^\prime, y^\prime, z^\prime$), as polymer gyroid templates yield preferential growth along the $[110]$ direction.

We performed finite-element micromagnetic simulations to study the magnetic-field-driven response of a nickel gyroid structure, using the software \texttt{finmag} \cite{2018finmag}.
Magnetic properties of nickel were described by a saturation magnetisation $M_\text{sat}=\SI{485}{kA/m}$ and an exchange constant $A_\text{ex}=\SI{8}{pJ/m}$. We assumed vanishing magnetocrystalline anisotropy, i.e., $K=0$.
The mesh of this gyroid simulation cell with a volume of $2\sqrt{2}a\times 2\sqrt{2}a\times 2a$, i.e., {\SI{184}{nm}$\times$\SI{184}{nm}$\times$\SI{130}{nm}}, as shown in Fig.~\ref{fig:gyroid_geometry}(a), was generated using COMSOL Multiphysics\textsuperscript{\textregistered} \cite{comsol} and contained 18196 nodes with a mean edge distance of \SI{4.4}{nm} (i.e., smaller than the magnetostatic exchange length $l_\text{ex}=\sqrt{2A_\text{ex}\,\mu_0^{-1}M_\text{sat}^{-2}}=\SI{7.5}{nm}$.

For applied magnetic fields $H\left(\sin\theta_H, 0, \cos\theta_H\right)$, equivalent to in-plane fields for gyroids grown by self-assembly, and a randomised initial spin configuration, micromagnetic configurations $\vec{m}(\vec{r})$ were obtained after relaxation.
The external field was then switched off, $H=0$, and the spin configuration again relaxed to obtain states at remanence. This process was repeated for angles $\theta_H$ between \ang{0} and \ang{360} in \ang{15} increments, and at field magnitudes $H=\SI{1}{T}$, \SI{100}{mT} and \SI{20}{mT}.

Analysis of the collective response was performed using the python package \texttt{networkx} \cite{networkx} by associating the local macrospins $\vec{s}_i$ of the gyroid structure with the edges of the underlying \textit{srs} network. This allowed to calculate properties such as the scalar spin chirality $\Omega_s$, local ice rules $A_\text{ice}$, as well as the anisotropic magnetoresistance as network resistance between specified nodes.

\section{Results and Discussion}

In the following, we will first discuss the global field-driven response of a single-gyroid structure, followed by assessing the local anisotropy of the individual struts to justify a macrospin picture of the gyroid network. We then turn to collective properties such as the magnetic order emerging on the triangular plaquettes and the global response from current transport through the network.

\subsection{Global Response}

\begin{figure}[tb]
	\centering
	\includegraphics[width=86mm]{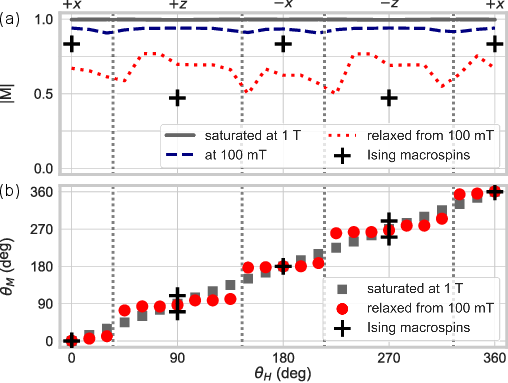}
	\caption{%
		\textbf{Global anisotropic response.}
		(a) Net magnetisation $|M|$ of gyroid dependent on field angle $\theta_H$.
		(b)~Angle of the mean magnetisation $\theta_M$ versus the field direction $\theta_H$.
		At high fields (\SI{1}{T}, gray line and gray dots) the magnetisation direction follows the field, i.e., $\theta_M=\theta_H$ and $M(\theta_H)\approx1$. 
		In the remanent state relaxed from high fields (from \SI{100}{mT}, shown with red dotted line and red squares) the magnetisation direction shows four distinct plateaus, showing preferential switching at angles related to the orientation of specific struts (vertical dotted lines).
	}
	\label{fig:global-magnetisation}
\end{figure}

From the field- and field-angle-dependent micromagnetic simulation, we obtain the magnetisation profile $\vec{m}(\vec{r})$ of the gyroid structure.
Figure~\ref{fig:global-magnetisation} shows the relation between the field angle $\theta_H$ in the $xz$ plane (with $\theta_H=0$ for $\vec{H}\parallel x$) and both the average magnetisation magnitude $|M|$ and direction $\theta_M(\theta_H)$ defined by 
	\begin{eqnarray}
		\tan(\theta_H) & = & \frac{H_z}{H_x} \, ,
		\\
		\tan\left(\theta_M(\theta_H)\right) & = &
		\frac{\left<\vec{m}(\vec{r}, \theta_H)\right>_z}{\left<\vec{m}(\vec{r}, \theta_H)\right>_x}
		\, .
	\end{eqnarray}
At high magnetic fields of \SI{1}{T} (solid gray line and gray squares) the sample magnetisation follows that of the applied field, i.e., $\theta_M=\theta_H$, indicating that the structure is saturated. In contrast, at fields of \SI{100}{mT} the sample shows a slightly non-isotropic response [dashed blue line in Fig.~\ref{fig:global-magnetisation}(a)].
Configurations relaxed from \SI{100}{mT} (dashed red line and red circles) and \SI{1}{T} (not shown) are qualitatively similar. They feature a reduced net moment and the magnetisation direction $\theta_M$ exhibits four distinct plateaus. The step-like reorientations occur around angles $\ang{0}\pm\alpha_s$ and $\ang{180}\pm\alpha_s$, marked by vertical dashed lines in Fig.~\ref{fig:global-magnetisation}, with angles $\alpha_s=\tan^{-1}(\sqrt{1/2})\approx\ang{35.4}$ in the $xy$ plane perpendicular to some of the struts.
The anisotropic global response and prominent demagnetisation therefore indicate that the gyroid network plays a major role to the hysteretic behaviour.

The response at \SI{20}{mT} and the obtained remanent state indicate unsystematic minor magnetic loops, and thus are excluded from the following discussion.

\subsection{Testing the Macrospin Assumption}

\begin{figure}[tb]
	\centering
	\includegraphics[width=86mm]{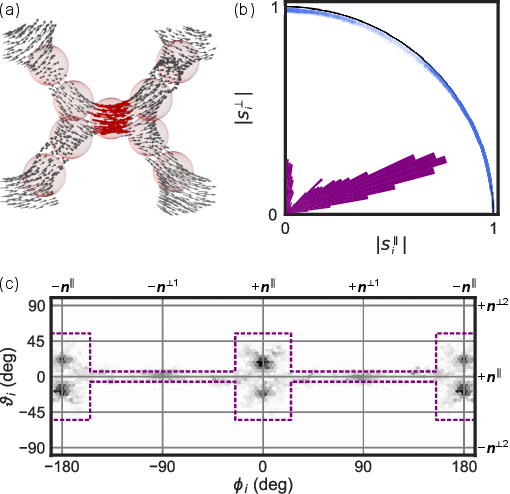}
	\caption[Macrospin assumption]{%
		\textbf{Macrospin assumption.} 
		(a)~The macrospin $\vec{s}_i$ is defined as the mean moment within non-overlapping volumes centred on each strut (red spheres).
		(b)~The magnitude of each strut moment $|\vec{s}_i|=|\vec{s}_i^\parallel+\vec{s}_i^\perp|$ at remanence (blue points) is close to one (black line), indicating that each strut acts as a macrospin. The rose diagram indicates non-Ising-like behaviour with a median inclination of about \ang{25} from the local direction $\hat{\vec{n}}_i^\parallel$.
		(c)~Moment orientation of the macrospins at remanence, described by spherical angles $\phi_i$ and $\vartheta_i$ in the local coordinate system, showing signatures of chiral magnetic order. In 90\% of the cases, the local moment lies within the region outlined in purple.
	}
	\label{fig:macrospin-assumption}
\end{figure}

\begin{figure*}[tb]
	\centering
	\includegraphics[width=178mm]{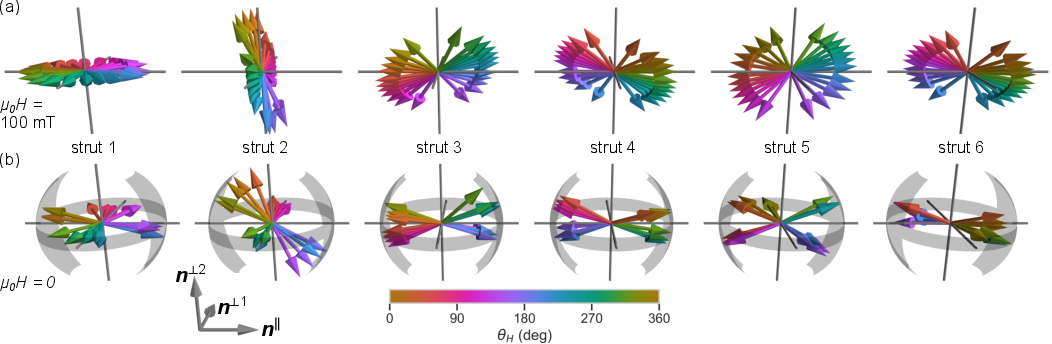}
	\caption[Local hysteretic behaviour.]{%
		\textbf{Local hysteretic behaviour}, with the strut moments in the local coordinate system $\hat{\mathbf{n}}_i^\parallel$, $\hat{\mathbf{n}}_i^{\perp 1}$, $\hat{\mathbf{n}}_i^{\perp 2}$ at (a)~\SI{100}{mT} and (b)~at remanence relaxed from \SI{100}{mT}. The coloured arrows denote the mean magnetisation corresponding to a specific field direction $\theta_H$ (colour bar).
		(a)~At \SI{100}{mT}, the moment mainly follows the direction of the applied field.
		(b)~At remanence, for each strut $i$ the mean moment relaxes from initial field direction $\theta_H$ into four angular quadrants, related to the four plateaus seen in the global response in Fig.~\ref{fig:global-magnetisation}(b).
	}
	\label{fig:strut-local-coords}
\end{figure*}

The simplest model to describe the magnetic order of a gyroid network would be an Ising system: As shown in Fig.~\ref{fig:global-magnetisation}(a), the idea is to average the moments in each strut to a macrospin $\vec{s}_i$, and that the shape anisotropy forces this macrospin to be parallel to the local strut direction $\hat{\mathbf{n}}^\parallel_i$. 
With this Ising macrospin assumption, values of $|M|$ and $\theta_M$ at remanence can be predicted at specific field angles $\theta_H$ by averaging over the six strut directions (Tab.~\ref{tab:strut-directions}), as shown by black crosses in Fig.~\ref{fig:global-magnetisation}. For the moment direction $\theta_M$ in Fig.~\ref{fig:global-magnetisation}(b), the Ising macrospin picture yields at  $\theta_M=\ang{0}$ at $\theta_H=\ang{0}$, and $\theta_M=\alpha_\text{twist}$ or $\theta_M=\ang{180}-\alpha_\text{twist}$ at $\theta_H=\ang{90}$, in reasonably good agreement with the micromagnetic simulation results.

The reasonable similarity between global and Ising-like behaviour justifies a closer look on the local anisotropies of the strut magnetisation: We obtain the magnitude and direction of strut macrospins $\vec{s}_i$ as the average of the local moment $\vec{m}(\vec{r})$ within non-overlapping spherical volumes centred on each strut position $\vec{r}_i^\text{cen}$, as shown in Fig.~\ref{fig:macrospin-assumption}(a):
\begin{equation}
	\vec{s}_i = \sum_{|\vec{r} - \vec{r}_i^\text{cen}| \le r_\text{s}} \vec{m}(\vec{r}) \quad. 
\end{equation}
With radius $r_\text{s} = \frac{2}{5} d_\text{NN} = a/(5\sqrt{2})$ the integration volumes contain about {$67\pm 11$} mesh points.
Using this approach we simplify the full micromagnetic configuration with 18196 mesh nodes to 192 individual struts. We furthermore discard struts at the boundary of the simulation volume, and in the following consider the properties of 160 struts for each field magnitude and angle.

As shown in Fig.~\ref{fig:macrospin-assumption}(b), at remanence the strut moments do indeed behave like macrospins, albeit not with the expected Ising-like anisotropy: Here, we separate the each strut moment into parallel and perpendicular components with respect to the local main strut direction $\hat{\mathbf{n}}_i^\parallel$: $\vec{s}_i = \vec{s}_i^{\parallel} + \vec{s}_i^{\perp}$.
Using this decomposition, in Fig.~\ref{fig:macrospin-assumption}(b) we see that the net amplitude $|\vec{s}_i|$ (blue dots) is close to one (black line), indicating a locally saturated magnetisation, with a reduction of at most 3\%.
Therefore, we can conclude that the macrospin assumption holds well, which is not entirely surprising: Each strut has a volume corresponding to a cylinder with \SI{10}{nm} diameter and \SI{25}{nm} length, dimensions which are comparable to the exchange length $l_\text{ex}=\SI{7.5}{nm}$ and thus supporting quasi-uniform strut magnetisation without the formation of domain walls.

\subsection{Local Magnetic Anisotropy}

Even though the struts' behaviour can be approximated with quasi-uniform macrospins, the local anisotropy does not favour simple Ising-like behaviour. 
This is illustrated by the rose diagram in Fig.~\ref{fig:macrospin-assumption}(b) (purple bars) which indicates that the macrospins $\vec{s}_i$ have a median inclination of about \ang{25} to the main strut axis $\hat{\mathbf{n}}_i^\parallel$.
Further insight to the local anisotropy can be gained by considering the spherical angles $\phi_i$ and $\vartheta_i$ denoting the macrospin orientation with respect to the local coordinate system defined in Tab.~\ref{tab:strut-directions}:
\begin{eqnarray}
	\tan(\phi_i) & = & \frac{\vec{s}_i\cdot\hat{\mathbf{n}}_i^{\perp1}}{\vec{s}_i\cdot\hat{\mathbf{n}}_i^{\parallel}} ,
	\label{eq:macrospin_phi}
	\\
	\tan(\vartheta_i) & = & \frac{ \vec{s}_i\cdot\hat{\mathbf{n}}_i^{\perp2} }{ 
		\sqrt{ (\vec{s}_i\cdot\hat{\mathbf{n}}_i^{\parallel})^2 + (\vec{s}_i\cdot\hat{\mathbf{n}}_i^{\perp1})^2 } 
		} .
		\label{eq:macrospin_theta}
\end{eqnarray}
As shown in Fig.~\ref{fig:macrospin-assumption}(c), 90\% of the moments fall into the area outlined by the dashed purple line. This permissible angular range, describing preferential anisotropy, combines "wings" centred at $\phi_i=\ang{0}$ and \ang{180} with $\Delta\phi_i\approx\pm\ang{25}$ and $\Delta\vartheta_i=\pm(\ang{90}-\alpha_s)$ with a "ring" in the $\hat{\vec{n}}_i^\parallel$--$\hat{\vec{n}}_i^{\perp1}$ plane ($\phi_i=\ang{-180}\ldots\ang{+180}$, $\Delta\vartheta_i=\pm\psi_\text{gDMI}$). 
Here, the angle $\psi_\text{gDMI}^\text{max}=\ang{6.5}$ is the maximum spin canting due to geometrical DMI \cite{2018Volkov} predicted for the tightest possible helix path [yellow line in Fig.~\ref{fig:gyroid_geometry}(c)].

The peculiar non-Ising, non-Heisenberg anisotropy is a direct consequence of at least three effects:
(1)~The wings in the anisotropy directly originates from the connectedness between vertices and the dominant exchange interaction which enforces magnetic continuity. As shown in Fig.~\ref{fig:gyroid_geometry}(b), neighbouring triangular plaquettes are non-coplanar, therefore tilting of the moment away from $\hat{\vec{n}}_i^\parallel$ can be energetically favourable to minimise the dipolar interactions between the net moments of the two vertices. However, tilting towards $\hat{\mathbf{n}}_i^{\perp 2}$ by more than $\pm(\ang{90}-\alpha_s)$ leads to energy-costly magnetic charges.
(2)~The ring corrresponds to small spin canting up to maximum values $\psi_\text{gDMI}$ due to geometrical DMI, showing a small asymmetry for fields at $\theta_H=\ang{-90}$ and $\theta{+90}$.
(3)~Slight asymmetries in the average angles $\left<\vartheta_i\right>$ of macrospins at \ang{0} and $\ang{\pm180}$ indicate a chiral contribution to the magnetic anisotropy, likely related to the underlying chiral right-handed crystal structure.

\subsection{Hysteretic Behaviour}

Figure~\ref{fig:strut-local-coords} gives further insight to the relationship between local anisotropy and hysteretic behaviour, broken down for individual struts $i$: Here, the local macrospin moments are indicated by black points, projected onto the corresponding coordinate system $\hat{\mathbf{n}}_i^\parallel$, $\hat{\mathbf{n}}_i^{\perp 1}$, and $\hat{\mathbf{n}}_i^{\perp 2}$. Coloured arrows indicate the mean moment direction at a given field angle $\theta_H$, averaged over all equivalent struts.

Figure~\ref{fig:strut-local-coords}(a) shows the behaviour at applied fields of \SI{0.1}{T}. As the field angle $\theta_H$ is defined in the global $xz$-plane, the struts are differently oriented with respect to the simulation coordinate frame. Therefore, the same field pulls the strut moments in significantly different directions with respect to the local coordinate system. A field magnitude of \SI{100}{mT} does not yet fully saturate the gyroid magnetisation, and the effect of local anisotropy is evident by the non-uniform rotation of the moment with the field direction.

Figure~\ref{fig:strut-local-coords}(b) shows the equivalent moment configurations at remanence relaxed from \SI{100}{mT}. The high-field state clearly influences the final configuration, and generally, four specific low-energy moment configurations can be classified for each strut, in accordance with the four plateaus of global magnetisation direction shown in Fig.~\ref{fig:global-magnetisation} (red circles).
The complex response of the local macrospins in combination with the highly-connected network therefore lead to emergent collective order in extended gyroid networks. 

\subsection{Spin Ice Rules}

\begin{figure}[t]
	\centering
	\includegraphics[width=86mm]{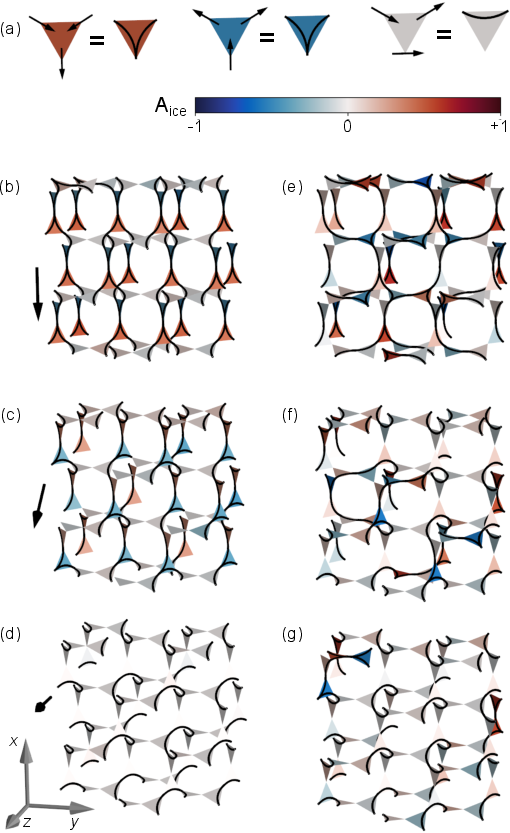}
	\caption[Lattice Flux]{%
		\textbf{Lattice flux in the gyroid network}, which can be represented by (a)~black lines that mark pair-wise in-out macrospin configurations. The colour scale for $A_\text{ice}$ denotes two-in-one-out (red), one-in-two-out (blue) and one-in-one-out (gray) triangular plaquettes.
		(b-d)~Macrospin flux lattices formed by application of magnetic fields $|H|=\SI{100}{mT}$s at angles (b)~$\theta_H=\ang{180}$, (c)~$\theta_H=\ang{135}$, and (d)~$\theta_H=\ang{90}$ (see black arrow).
		(e-g)~At remanence, relaxed from the respective configuration shown on the left, the overall magnetic connectivity and the 3d character of the flux lines through the gyroid lattice has increased.
	}
	\label{fig:ASI-network}
\end{figure}

To discuss possible collective effects in the gyroid network, we first will focus on the statistical properties and magnetic connectivity of the corner-sharing triangular plaquettes, as shown in Fig.~\ref{fig:gyroid_geometry}(c). 
%
In many cases, the spin order of frustrated magnetic networks with triangular plaquettes is governed by \textit{local} ice rules instead of long-ranged \textit{global} order \cite{2004Diep, Lacroix11, 2012Branford, 2016Canals, 2017Gartside}.
%
To discuss the concept of ice rules in the gyroid network, in the following we will consider the scalar Ising-like component of the strut macrospin $s_i^I$ defined by 
\begin{equation}
	s_i^I = 
		\frac{ \vec{s}_{i} \cdot \vec{r}^{\parallel, \text{in}} }{ |\vec{r}^{\parallel, \text{in}}| }
		\quad \text{with} \quad 
		\vec{r}^{\parallel, \text{in}} = \vec{r}_i-\left<\vec{r}\right>_{1,2,3}
\end{equation}
Here, $\left<\vec{r}\right>_{1,2,3}$ denotes the centre of the corresponding triangular plaquette, as the mean value of coordinates from struts $1$, $2$, and $3$. Thus, for Ising-like macrospins the values of $s_i^I$ will be $+1$ ($-1$) for moments pointing into (out from) the centre of the corresponding plaquette.

First, to test for ice-like correlations we consider the scalar spin chirality \cite{2001Taguchi, 2008Park} of each vertex as 
\begin{equation}
	\Omega_s = \frac{1}{3} \left( s_1^Is_2^I + s_2^Is_3^I + s_3^Is_1^I \right) 
	\, .
	\label{eq:spin_chirality}
\end{equation}
The scalar spin chirality $\Omega_s$ can take two limiting values, $\Omega_s=+1$, corresponding to all-in or all-out moment configurations, and $\Omega_s=-1/3$, which quantifies local ice-like two-in-one-out (or vice versa) configurations.
We found no single case corresponding an all-in or all-out moment configuration, as such monopole-like configurations are too energetically costly to occur in our exchange-dominated nanoscale magnetic gyroid structure. 
At remanence, the median value for the spin chirality is $\left<\Omega_s\right>=\num{-0.29}$, i.e., close to the theoretical value of $-1/3$, independent of the initial field direction $\theta_H$. This finding indicates that the local magnetic order in gyroid networks is governed by spin-ice rules. 

To further quantify the local magnetic order, we now calculate the sum of the Ising-like moments of each triangular plaquette combining struts $1$, $2$, and $3$:
\begin{equation}
	A_\text{ice} = s_1^I + s_2^I + s_3^I
	\quad .
	\label{eq:ice_rule}
\end{equation}
The quantity $A_\text{ice}$ allows to easily distinguish between two-in-one-out ($A_\text{ice}=+1$) and one-in-two-out ($A_\text{ice}=-1$) moment configurations. Since the local moments do not strictly follow the in-or-out Ising anisotropy, i.e., $|s_i^I|\leq 1$,  values of $A_\text{ice}$ between these two limits are also allowed. 
In particular, we find that values of $A_\text{ice}$ around zero are highly likely, corresponding to triangular plaquettes where one moment is approximately parallel to the one-in-one-out magnetisation of the two opposite struts [Fig.~\ref{fig:ASI-network}(a)].
Because of the twist between neighbouring plaquettes, this non-Ising moment will also be \textit{non-coplanar} with respect to at least one triangle plane, which enables non-zero vector chirality terms \cite{2008Park} related to additional magnetic properties, such as non-reciprocal magnetotransport and spin-wave propagation \cite{2016Seki, 2018Tokura, 2020Barman}.

Because of the dominant exchange interaction, the magnetic connection formed by a single strut is continuous, i.e., a macrospin which points \textit{out} of a triangular plaquette must point \textit{into} the neighbouring one.
This continuity allows to define a magnetic flux across the three-dimensional gyroid lattice (or \textit{srs} net), where each triangle carries at least one one-in-one-out flux line, or allows the bifurcation of two flux lines via two-in-one-out or one-in-two-out moment configurations.

Fig.~\ref{fig:ASI-network} illustrates the 3D connectivity of the magnetic flux  at different fields and the corresponding remanent states. 
As shown in Fig.~\ref{fig:ASI-network}(a), the magnetic order is simplified by using triangular plaquettes coloured by the value of $A_\text{ice}$. Neighbouring corners $i$ and $j$ are connected by a black ``flux'' line if the pair corresponds to an one-in-one-out moment configuration, i.e., $s_i^I s_j^I <0$, with the additional condition that the two moments are sufficiently Ising-like, i.e., $|s^I_{i,j}|\geq0.5$.
Under applied field, Fig.~\ref{fig:ASI-network}(b-d), the flux forms a regular pattern through the gyroid lattice, as the moments are largely aligned parallel to the field. Depending on the field direction $\theta_H$, the flux distribution is mostly confined to the $xz$ plane [(b,d) for $\theta_H=\ang{180}$ or \ang{135}, respectively], or exhibits one-dimensional flux channels along $z$ [(c), $\theta_H=\ang{90}$].
The magnetic configuration at remanence shown in Fig.~\ref{fig:ASI-network}(e-g), show a higher connectivity and more plaquettes with ice-like two-in/two-out configuration compared to the high-field states they were relaxed from. 
This is especially the case of $\theta_H=\ang{0}$ [Fig.~\ref{fig:ASI-network}(e)], which features a complex three-dimensional flux network. Regardless of the increased magnetic connectivity, however, many triangular plaquettes still feature one flux line only, due to a perpendicular moment on the third macrospin.
Simulation results from minor loops (relaxed from \SI{20}{mT}, not shown) resulted in a higher ratio of ice-like correlations, indicating that a suitable demagnetisation protocol could be used to relax the gyroid lattice to a low-energy configuration with predominant ice-like correlations indicative of a highly-frustrated spin system.

\subsection{Magnetotransport}

\begin{figure}[t!]
	\centering
	\includegraphics[width=86mm]{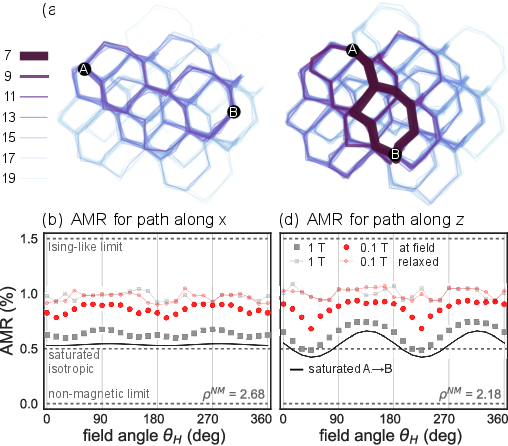}
	\caption[Anisotropic magnetoresistance]{%
		\textbf{Anisotropic magnetoresistance} through a finite-size gyroid network.
		(a)~Possible paths $A\rightarrow B$ connecting nodes $A$ and $B$ in the $x$ direction, with different length indicated by line colour and width.
		(b)~Dependence of relative AMR signal between nodes $A$ and $B$ on field angle $\theta_H$. Solid and dashed lines denote limiting cases to the magnetoresistance in the gyroid network, squares and dots show the AMR signal calculated from the macrospin analysis of the micromagnetic simulations.
		$\rho_\text{NM}$ denotes the non-magnetic network resistance between $A$ and $B$. 
		(c,d)~Equivalent observations for paths between $A\rightarrow B$ along the $z$ direction, exhibiting more pronounced angular variation with $\theta_H$.
	}
	\label{fig:AMR}
\end{figure}

Finally, we consider the complex directional magnetotransport signatures emerging in the gyroid, which can be used to fingerprint the magnetic order and {local} anisotropy \cite{2016Seki, 2018Tokura}.
Because of the finite-size {volume} of the micromagnetic simulations, we here discuss the most salient features of the anisotropic magneto\-resistance (AMR) only, using a simplified geometrical model for magnetoresistance in networks \cite{2005Wang}. 

For each strut in the gyroid lattice the AMR results in a variation of the local longitudinal resistance $\rho_i(\varphi_i)$ in dependence of the angle $\varphi_i=\sphericalangle\left(\vec{j}_i,\vec{s}_i\right)$ between the charge current flow direction $\vec{j}_i\parallel\hat{\vec{n}}_\parallel$ and the macrospin $\vec{s}_i$, with
\begin{equation}
	\rho_i(\varphi_i) = \rho_0 \left( 1 + \Delta_\text{AMR} \cos^2(\varphi_i) \right)
	\, ,
	\label{eq:AMR}
\end{equation}
where $\rho_0$ and $\Delta_\text{AMR}$ are the non-magnetic resistance and the relative magnitude of the AMR effect of the underlying bulk material, respectively.
For convenience, we set $\rho_0=1$.
For nickel nanowires the typical AMR magnitude is in the order of $\Delta_\text{AMR}=\num{1.5}\%$ \cite{2000Pignard}.

There are a multitude of possible paths connecting two chosen nodes $A$ and $B$ through the gyroid network, which act as parallel conduction channels with piece-wise local resistances $\rho_i(\varphi_i)$. By applying Kirchhoffs law, one can calculate the effective network resistance $\rho_{A\rightarrow B}$, here using the function \texttt{resistance\_distance} of the python package \texttt{networkx} \cite{1993Klein, networkx}. As the magnetic order is highly-dependent on the field direction, we thus expect significant variation of $\rho^\text{AMR}_{A\rightarrow B}(\theta_H)$ with both the field magnitude $H$ and angle $\theta_H$ as well as the choice of $A$ and $B$.

Figure~\ref{fig:AMR} shows the AMR response for two node pairs $A$ and $B$ separated along the $x$ [Fig.~\ref{fig:AMR}(a,b)] and along the $z$ direction [Fig.~\ref{fig:AMR}(c,d)]. 
The top graphs (a,c) illustrate parallel connections $A\rightarrow B$, highlighting the exponentially increasing number of possible paths with increasing path length.
The bottom graphs (b, d) show the angular dependence of the AMR calculated from the micromagnetic simulations, which originates from a mixture of the local magnetic anisotropy and the multitude of parallel paths through the network.

Before turning to the macrospin results, we briefly discuss three general limiting cases to the AMR within the gyroid network indicated by dashed gray lines in Figs.~\ref{fig:AMR}(b,d):
First, for the minimum non-magnetic limit (AMR=0\%, bottom line) the network resistance $\rho_{A\rightarrow B}^\text{NM}$ is increased compared to the bulk value $\rho_0=1$, but significantly smaller than the length of the shortest paths $A\rightarrow B$ would imply, with $\rho^\text{NM}=\num{2.68}$ but $l_\text{min}=\num{11}d_\text{NN}$ for (a,b), and $\rho^\text{NM}=\num{2.18}$ but $l_\text{min}=\num{7}d_\text{NN}$ for (c,d). 
Second, the maximum limit for AMR (1.5\%, top line) is achieved for perfect Ising-like macrospins, with $\vec{s}_i\parallel\hat{\vec{n}}_\parallel\parallel\vec{j}_i$. 
Third, for an infinite-size gyroid lattice the AMR response between nodes $A\rightarrow B$ is isotropic with respect to $\theta_H$, at $\text{AMR}_\text{gyroid}^\text{isotropic}=\frac{1}{3}\max(\text{AMR}_\text{bulk})$, here AMR=0.5\% (middle dashed line).

The field-saturated case highlights one stark difference between bulk-like AMR and AMR in a gyroid network: In bulk, the maximum AMR is observed at saturation with $\vec{M}\parallel\vec{H}$, whereas in the gyroid network, high fields destroy any Ising-like state that leads to maximum AMR. 
Instead, as shown by the black lines in Fig.~\ref{fig:AMR}(b,d), the field-saturated AMR, with $\vec{s}_i\parallel\vec{H}$ for each strut $i$ along paths $A\rightarrow B$, lies in-between the previously discussed limiting cases. The angular variation with $\theta_H$ is a direct consequence of the finite-size and the relative importance of different strut directions within the connecting paths, as illustrated by the higher anisotropy in Fig.~\ref{fig:AMR}(d) compared to (b). 
This behaviour means that measurements at saturation, rather than the magnetoresistance at remanence, are more likely to privide a reliable normalisation in experimental studies of the magnetoresistive response of gyroid networks.

Somewhat surprisingly, the AMR at \SI{1}{T} [gray squares in Figs.~\ref{fig:AMR}(b,d)] lies above the predicted saturated response (black line) and is also more anisotropic. This deviation is due to the fact that even though the net magnetisation seems saturated, see Fig.~\ref{fig:global-magnetisation}(a), the macrospins $\vec{s}_i$ still can have a slight inclination to the direction of $\vec{H}$ of about \ang{3.5} to \ang{5}. This value is consistent with analytical predictions for spin canting induced by geometry-induced DMI ($\psi_\text{gDMI}^\text{max}=\ang{6.5}$) discussed above. 

As the field magnitude is decreased to \SI{100}{mT} [red circles in Fig.~\ref{fig:AMR}(b,d)] the net AMR signal increases, and its angular variation is modified. This observation can be related to both the non-monotonic rotation of the individual macrospins due to the intrinsic local anisotropy as represented in Fig.~\ref{fig:strut-local-coords}(a), as well as the collective response related to the emergent flux lattice shown in Fig.~\ref{fig:ASI-network}.

Finally, the AMR signal at remanence (lighter dots) lies about halfway between the limits of isotropic field-saturation and the Ising-like limit, in agreement with the local non-Ising anisotropy. The angular variation of the remanent AMR is rather weak.

In conclusion, magnetoresistance signatures of gyroids give insight to the effective local magnetic anisotropy and emergent collective behaviour. In comparison to 2D artificial spin systems \cite{2005Wang, 2012Branford, 2017Le}, there are notable differences in the magnetoresistive response of gyroids: 
First, there are many more possible conduction pathways, as wires can cross in 3D geometries but not in planar devices. 
Second, due to the regular 3D arrangement of struts, the spin order is inherently non-collinear \textit{and} non-coplanar, irrespective of the direction and magnitude of any applied magnetic field. 
In combination with the high degree of frustration, magnetoresistance measurements including the anomalous Hall effect \cite{2012Branford, 2017Le} and chiral magnetoresistance and non-reciprocal spin-wave propagation due to non-vanishing vector spin chirality \cite{2001Taguchi, 2016Seki, 2018Tokura}  therefore could be the ideal tool to elucidate the emerging collective response of magnetic gyroids.

\section{Conclusions and Outlook}
\label{sec:conclusions}

In this work we considered the complex spin order of a gyroid networks at applied magnetic fields and at remanence.
Using micromagnetic simulations, we reveal that for nanoscale nickel gyroids the individual struts can be described as quasi-uniform macrospins. Their complex configuration is affected by both the three-dimensional network connectivity as well as modified by a effective chiral exchange term, respectively geometrical DMI.
While the gyroid network is built from corner-sharing triangular plaquettes, and thus is a prime host for geometrically-frustated spin order, the deviation from local Ising-like magnetic anisotropy reduce ice-like correlations. 

We find that magnetotransport signatures reflect the complexity of spin order within the gyroid lattice and are different to magnetoresistive behaviour of both bulk samples and 2D artificial spin systems. 
Especially in comparison to planar devices, the 3D geometry and connectivity, truly 3D spin order in response to 3D fields, and the multitude of parallel conduction channels of the regular gyroid network result in an extensive manifold of magnetic states as well as gives many possible choices for magnetotransport geometries. This vast phase space therefore is ideal to be explored for future three-dimensional spintronic applications \cite{2017Fernandez, 2022Bhattacharya}.

For future experimental exploration of nanoscale gyroids prepared by different growth methods, such as polymer self-assembly {\cite{2015Dolan, 2015Wu_a, 2018Dolan}, focused-electron beam deposition (FEBID) \cite{2020FernandezPacheco, 2020Skoric}, or two-photon nanolithography \cite{2018Williams, 2022Pip, 2023Berg} we identify two main aspects of emergent magnetic order to explore: 
First, our results indicate that highly-frustrated spin configurations can be prepared with suitable demagnetisation protocols in gyroid networks with enhanced Ising-like macrospin behaviour (e.g., stabilised by choice of materials or by preparing networks with larger lattice constants), and likely lead to interesting collective 3D artificial spin ice behaviour. 
Second, due to their inherent non-coplanar spin order and thus non-trivial vector spin chirality, gyroid networks are ideal candidates to host directional magnetotransport and non-reciprocal spin wave propagation \cite{2001Taguchi, 2016Seki, 2018Tokura}.
Such emergent properties and the intrinsic stochasticity related to frustated magnetic order in nanoscale gyroids therefore makes them a rich platform to investigate 3D spintronic networks for probabilistic and neuromorphic computing \cite{2018Fukami, 2021Dawidek, 2022Gartside, 2022Bhattacharya}. 

~\linebreak
The raw data supporting the findings of this study are openly available at the zenodo repository \cite{data_repository}.

\begin{acknowledgments}
	
	The authors thank Attila Kakay and Amalio {Fernández}-Pacheco for helpful discussions.
	N.L.\ received funding from the European Research Council (ERC) under European Union's Horizon 2020 research and innovation programme under the Marie Sklodowska Curie Grant Agreement No.~844304 (LICONAMCO), as well as support by the European Community under the Horizon 2020 Program, Contract No.~101001290 (3DNANOMAG).
	The work of A.K.\ was financially supported by JSPS KAKENHI Grant No.~18J20396.
	The work of J.L.\ and S.F.\ was supported by JSPS KAKENHI 21K04816 and 19H05622 and the Graduate Program for Spintronics (GP-Spin) as well as Cooperative Research Projects of CSIS and CSRN, Tohoku University.

\end{acknowledgments}


%


\begin{thebibliography}{66}%
\makeatletter
\providecommand \@ifxundefined [1]{%
 \@ifx{#1\undefined}
}%
\providecommand \@ifnum [1]{%
 \ifnum #1\expandafter \@firstoftwo
 \else \expandafter \@secondoftwo
 \fi
}%
\providecommand \@ifx [1]{%
 \ifx #1\expandafter \@firstoftwo
 \else \expandafter \@secondoftwo
 \fi
}%
\providecommand \natexlab [1]{#1}%
\providecommand \enquote  [1]{``#1''}%
\providecommand \bibnamefont  [1]{#1}%
\providecommand \bibfnamefont [1]{#1}%
\providecommand \citenamefont [1]{#1}%
\providecommand \href@noop [0]{\@secondoftwo}%
\providecommand \href [0]{\begingroup \@sanitize@url \@href}%
\providecommand \@href[1]{\@@startlink{#1}\@@href}%
\providecommand \@@href[1]{\endgroup#1\@@endlink}%
\providecommand \@sanitize@url [0]{\catcode `\\12\catcode `\$12\catcode
  `\&12\catcode `\#12\catcode `\^12\catcode `\_12\catcode `\%12\relax}%
\providecommand \@@startlink[1]{}%
\providecommand \@@endlink[0]{}%
\providecommand \url  [0]{\begingroup\@sanitize@url \@url }%
\providecommand \@url [1]{\endgroup\@href {#1}{\urlprefix }}%
\providecommand \urlprefix  [0]{URL }%
\providecommand \Eprint [0]{\href }%
\providecommand \doibase [0]{https://doi.org/}%
\providecommand \selectlanguage [0]{\@gobble}%
\providecommand \bibinfo  [0]{\@secondoftwo}%
\providecommand \bibfield  [0]{\@secondoftwo}%
\providecommand \translation [1]{[#1]}%
\providecommand \BibitemOpen [0]{}%
\providecommand \bibitemStop [0]{}%
\providecommand \bibitemNoStop [0]{.\EOS\space}%
\providecommand \EOS [0]{\spacefactor3000\relax}%
\providecommand \BibitemShut  [1]{\csname bibitem#1\endcsname}%
\let\auto@bib@innerbib\@empty
\bibitem [{\citenamefont {Skjærvø}\ \emph {et~al.}(2019)\citenamefont
  {Skjærvø}, \citenamefont {Marrows}, \citenamefont {Stamps},\ and\
  \citenamefont {Heyderman}}]{2019Skjaervoe}%
  \BibitemOpen
  \bibfield  {author} {\bibinfo {author} {\bibfnamefont {S.~H.}\ \bibnamefont
  {Skjærvø}}, \bibinfo {author} {\bibfnamefont {C.~H.}\ \bibnamefont
  {Marrows}}, \bibinfo {author} {\bibfnamefont {R.~L.}\ \bibnamefont
  {Stamps}},\ and\ \bibinfo {author} {\bibfnamefont {L.~J.}\ \bibnamefont
  {Heyderman}},\ }\bibfield  {title} {\bibinfo {title} {Advances in artificial
  spin ice},\ }\href {https://doi.org/10.1038/s42254-019-0118-3} {\bibfield
  {journal} {\bibinfo  {journal} {Nature Reviews Physics}\ }\textbf {\bibinfo
  {volume} {2}},\ \bibinfo {pages} {13} (\bibinfo {year} {2019})}\BibitemShut
  {NoStop}%
\bibitem [{\citenamefont {Gliga}\ \emph {et~al.}(2020)\citenamefont {Gliga},
  \citenamefont {Iacocca},\ and\ \citenamefont {Heinonen}}]{2020Gliga}%
  \BibitemOpen
  \bibfield  {author} {\bibinfo {author} {\bibfnamefont {S.}~\bibnamefont
  {Gliga}}, \bibinfo {author} {\bibfnamefont {E.}~\bibnamefont {Iacocca}},\
  and\ \bibinfo {author} {\bibfnamefont {O.~G.}\ \bibnamefont {Heinonen}},\
  }\bibfield  {title} {\bibinfo {title} {Dynamics of reconfigurable artificial
  spin ice: Toward magnonic functional materials},\ }\href
  {https://doi.org/10.1063/1.5142705} {\bibfield  {journal} {\bibinfo
  {journal} {APL Materials}\ }\textbf {\bibinfo {volume} {8}},\ \bibinfo
  {pages} {040911} (\bibinfo {year} {2020})}\BibitemShut {NoStop}%
\bibitem [{\citenamefont {Barman}\ \emph {et~al.}(2021)\citenamefont {Barman},
  \citenamefont {Gubbiotti}, \citenamefont {Ladak}, \citenamefont {Adeyeye},
  \citenamefont {Krawczyk}, \citenamefont {Gräfe}, \citenamefont {Adelmann},
  \citenamefont {Cotofana}, \citenamefont {Naeemi}, \citenamefont {Vasyuchka},
  \citenamefont {Hillebrands}, \citenamefont {Nikitov}, \citenamefont {Yu},
  \citenamefont {Grundler}, \citenamefont {Sadovnikov}, \citenamefont
  {Grachev}, \citenamefont {Sheshukova}, \citenamefont {Duquesne},
  \citenamefont {Marangolo}, \citenamefont {Gyorgy}, \citenamefont {Porod},
  \citenamefont {Demidov}, \citenamefont {Urazhdin}, \citenamefont
  {Demokritov}, \citenamefont {Albisetti}, \citenamefont {Petti}, \citenamefont
  {Bertacco}, \citenamefont {Schulteiss}, \citenamefont {Kruglyak},
  \citenamefont {Poimanov}, \citenamefont {Sahoo}, \citenamefont {Sinha},
  \citenamefont {Yang}, \citenamefont {Muenzenberg}, \citenamefont {Moriyama},
  \citenamefont {Mizukami}, \citenamefont {Landeros}, \citenamefont {Gallardo},
  \citenamefont {Carlotti}, \citenamefont {Kim}, \citenamefont {Stamps},
  \citenamefont {Camley}, \citenamefont {Rana}, \citenamefont {Otani},
  \citenamefont {Yu}, \citenamefont {Yu}, \citenamefont {Bauer}, \citenamefont
  {Back}, \citenamefont {Uhrig}, \citenamefont {Dobrovolskiy}, \citenamefont
  {van Dijken}, \citenamefont {Budinska}, \citenamefont {Qin}, \citenamefont
  {Chumak}, \citenamefont {Khitun}, \citenamefont {Nikonov}, \citenamefont
  {Young}, \citenamefont {Zingsem},\ and\ \citenamefont
  {Winklhofer}}]{2021Barman}%
  \BibitemOpen
  \bibfield  {author} {\bibinfo {author} {\bibfnamefont {A.}~\bibnamefont
  {Barman}}, \bibinfo {author} {\bibfnamefont {G.}~\bibnamefont {Gubbiotti}},
  \bibinfo {author} {\bibfnamefont {S.}~\bibnamefont {Ladak}}, \bibinfo
  {author} {\bibfnamefont {A.~O.}\ \bibnamefont {Adeyeye}}, \bibinfo {author}
  {\bibfnamefont {M.}~\bibnamefont {Krawczyk}}, \bibinfo {author}
  {\bibfnamefont {J.}~\bibnamefont {Gräfe}}, \bibinfo {author} {\bibfnamefont
  {C.}~\bibnamefont {Adelmann}}, \bibinfo {author} {\bibfnamefont
  {S.}~\bibnamefont {Cotofana}}, \bibinfo {author} {\bibfnamefont
  {A.}~\bibnamefont {Naeemi}}, \bibinfo {author} {\bibfnamefont {V.~I.}\
  \bibnamefont {Vasyuchka}}, \bibinfo {author} {\bibfnamefont {B.}~\bibnamefont
  {Hillebrands}}, \bibinfo {author} {\bibfnamefont {S.~A.}\ \bibnamefont
  {Nikitov}}, \bibinfo {author} {\bibfnamefont {H.}~\bibnamefont {Yu}},
  \bibinfo {author} {\bibfnamefont {D.}~\bibnamefont {Grundler}}, \bibinfo
  {author} {\bibfnamefont {A.}~\bibnamefont {Sadovnikov}}, \bibinfo {author}
  {\bibfnamefont {A.~A.}\ \bibnamefont {Grachev}}, \bibinfo {author}
  {\bibfnamefont {S.~E.}\ \bibnamefont {Sheshukova}}, \bibinfo {author}
  {\bibfnamefont {J.-Y.}\ \bibnamefont {Duquesne}}, \bibinfo {author}
  {\bibfnamefont {M.}~\bibnamefont {Marangolo}}, \bibinfo {author}
  {\bibfnamefont {C.}~\bibnamefont {Gyorgy}}, \bibinfo {author} {\bibfnamefont
  {W.}~\bibnamefont {Porod}}, \bibinfo {author} {\bibfnamefont {V.~E.}\
  \bibnamefont {Demidov}}, \bibinfo {author} {\bibfnamefont {S.}~\bibnamefont
  {Urazhdin}}, \bibinfo {author} {\bibfnamefont {S.}~\bibnamefont
  {Demokritov}}, \bibinfo {author} {\bibfnamefont {E.}~\bibnamefont
  {Albisetti}}, \bibinfo {author} {\bibfnamefont {D.}~\bibnamefont {Petti}},
  \bibinfo {author} {\bibfnamefont {R.}~\bibnamefont {Bertacco}}, \bibinfo
  {author} {\bibfnamefont {H.}~\bibnamefont {Schulteiss}}, \bibinfo {author}
  {\bibfnamefont {V.~V.}\ \bibnamefont {Kruglyak}}, \bibinfo {author}
  {\bibfnamefont {V.~D.}\ \bibnamefont {Poimanov}}, \bibinfo {author}
  {\bibfnamefont {A.~K.}\ \bibnamefont {Sahoo}}, \bibinfo {author}
  {\bibfnamefont {J.}~\bibnamefont {Sinha}}, \bibinfo {author} {\bibfnamefont
  {H.}~\bibnamefont {Yang}}, \bibinfo {author} {\bibfnamefont {M.}~\bibnamefont
  {Muenzenberg}}, \bibinfo {author} {\bibfnamefont {T.}~\bibnamefont
  {Moriyama}}, \bibinfo {author} {\bibfnamefont {S.}~\bibnamefont {Mizukami}},
  \bibinfo {author} {\bibfnamefont {P.}~\bibnamefont {Landeros}}, \bibinfo
  {author} {\bibfnamefont {R.~A.}\ \bibnamefont {Gallardo}}, \bibinfo {author}
  {\bibfnamefont {G.}~\bibnamefont {Carlotti}}, \bibinfo {author}
  {\bibfnamefont {J.-V.}\ \bibnamefont {Kim}}, \bibinfo {author} {\bibfnamefont
  {R.~L.}\ \bibnamefont {Stamps}}, \bibinfo {author} {\bibfnamefont {R.~E.}\
  \bibnamefont {Camley}}, \bibinfo {author} {\bibfnamefont {B.}~\bibnamefont
  {Rana}}, \bibinfo {author} {\bibfnamefont {Y.}~\bibnamefont {Otani}},
  \bibinfo {author} {\bibfnamefont {W.}~\bibnamefont {Yu}}, \bibinfo {author}
  {\bibfnamefont {T.}~\bibnamefont {Yu}}, \bibinfo {author} {\bibfnamefont
  {G.~E.~W.}\ \bibnamefont {Bauer}}, \bibinfo {author} {\bibfnamefont {C.~H.}\
  \bibnamefont {Back}}, \bibinfo {author} {\bibfnamefont {G.~S.}\ \bibnamefont
  {Uhrig}}, \bibinfo {author} {\bibfnamefont {O.~V.}\ \bibnamefont
  {Dobrovolskiy}}, \bibinfo {author} {\bibfnamefont {S.}~\bibnamefont {van
  Dijken}}, \bibinfo {author} {\bibfnamefont {B.}~\bibnamefont {Budinska}},
  \bibinfo {author} {\bibfnamefont {H.}~\bibnamefont {Qin}}, \bibinfo {author}
  {\bibfnamefont {A.}~\bibnamefont {Chumak}}, \bibinfo {author} {\bibfnamefont
  {A.}~\bibnamefont {Khitun}}, \bibinfo {author} {\bibfnamefont {D.~E.}\
  \bibnamefont {Nikonov}}, \bibinfo {author} {\bibfnamefont {I.~A.}\
  \bibnamefont {Young}}, \bibinfo {author} {\bibfnamefont {B.}~\bibnamefont
  {Zingsem}},\ and\ \bibinfo {author} {\bibfnamefont {M.}~\bibnamefont
  {Winklhofer}},\ }\bibfield  {title} {\bibinfo {title} {The 2021 magnonics
  roadmap},\ }\href
  {http://iopscience.iop.org/article/10.1088/1361-648X/abec1a} {\bibfield
  {journal} {\bibinfo  {journal} {Journal of Physics: Condensed Matter}\
  }\textbf {\bibinfo {volume} {33}},\ \bibinfo {pages} {413001} (\bibinfo
  {year} {2021})}\BibitemShut {NoStop}%
\bibitem [{\citenamefont {Branford}\ \emph {et~al.}(2012)\citenamefont
  {Branford}, \citenamefont {Ladak}, \citenamefont {Read}, \citenamefont
  {Zeissler},\ and\ \citenamefont {Cohen}}]{2012Branford}%
  \BibitemOpen
  \bibfield  {author} {\bibinfo {author} {\bibfnamefont {W.~R.}\ \bibnamefont
  {Branford}}, \bibinfo {author} {\bibfnamefont {S.}~\bibnamefont {Ladak}},
  \bibinfo {author} {\bibfnamefont {D.~E.}\ \bibnamefont {Read}}, \bibinfo
  {author} {\bibfnamefont {K.}~\bibnamefont {Zeissler}},\ and\ \bibinfo
  {author} {\bibfnamefont {L.~F.}\ \bibnamefont {Cohen}},\ }\bibfield  {title}
  {\bibinfo {title} {Emerging chirality in artificial spin ice},\ }\href
  {https://doi.org/10.1126/science.1211379} {\bibfield  {journal} {\bibinfo
  {journal} {Science}\ }\textbf {\bibinfo {volume} {335}},\ \bibinfo {pages}
  {1597} (\bibinfo {year} {2012})}\BibitemShut {NoStop}%
\bibitem [{\citenamefont {Canals}\ \emph {et~al.}(2016)\citenamefont {Canals},
  \citenamefont {Chioar}, \citenamefont {Nguyen}, \citenamefont {Hehn},
  \citenamefont {Lacour}, \citenamefont {Montaigne}, \citenamefont {Locatelli},
  \citenamefont {Mentes}, \citenamefont {Burgos},\ and\ \citenamefont
  {Rougemaille}}]{2016Canals}%
  \BibitemOpen
  \bibfield  {author} {\bibinfo {author} {\bibfnamefont {B.}~\bibnamefont
  {Canals}}, \bibinfo {author} {\bibfnamefont {I.-A.}\ \bibnamefont {Chioar}},
  \bibinfo {author} {\bibfnamefont {V.-D.}\ \bibnamefont {Nguyen}}, \bibinfo
  {author} {\bibfnamefont {M.}~\bibnamefont {Hehn}}, \bibinfo {author}
  {\bibfnamefont {D.}~\bibnamefont {Lacour}}, \bibinfo {author} {\bibfnamefont
  {F.}~\bibnamefont {Montaigne}}, \bibinfo {author} {\bibfnamefont
  {A.}~\bibnamefont {Locatelli}}, \bibinfo {author} {\bibfnamefont {T.~O.}\
  \bibnamefont {Mentes}}, \bibinfo {author} {\bibfnamefont {B.~S.}\
  \bibnamefont {Burgos}},\ and\ \bibinfo {author} {\bibfnamefont
  {N.}~\bibnamefont {Rougemaille}},\ }\bibfield  {title} {\bibinfo {title}
  {Fragmentation of magnetism in artificial kagome dipolar spin ice},\ }\href
  {https://doi.org/10.1038/ncomms11446} {\bibfield  {journal} {\bibinfo
  {journal} {Nat Commun}\ }\textbf {\bibinfo {volume} {7}},\ \bibinfo {pages}
  {11446} (\bibinfo {year} {2016})}\BibitemShut {NoStop}%
\bibitem [{\citenamefont {Gartside}\ \emph {et~al.}(2018)\citenamefont
  {Gartside}, \citenamefont {Arroo}, \citenamefont {Burn}, \citenamefont
  {Bemmer}, \citenamefont {Moskalenko}, \citenamefont {Cohen},\ and\
  \citenamefont {Branford}}]{2017Gartside}%
  \BibitemOpen
  \bibfield  {author} {\bibinfo {author} {\bibfnamefont {J.~C.}\ \bibnamefont
  {Gartside}}, \bibinfo {author} {\bibfnamefont {D.~M.}\ \bibnamefont {Arroo}},
  \bibinfo {author} {\bibfnamefont {D.~M.}\ \bibnamefont {Burn}}, \bibinfo
  {author} {\bibfnamefont {V.~L.}\ \bibnamefont {Bemmer}}, \bibinfo {author}
  {\bibfnamefont {A.}~\bibnamefont {Moskalenko}}, \bibinfo {author}
  {\bibfnamefont {L.~F.}\ \bibnamefont {Cohen}},\ and\ \bibinfo {author}
  {\bibfnamefont {W.~R.}\ \bibnamefont {Branford}},\ }\bibfield  {title}
  {\bibinfo {title} {Realization of ground state in artificial kagome spin ice
  via topological defect-driven magnetic writing},\ }\href
  {https://doi.org/10.1038/s41565-017-0002-1} {\bibfield  {journal} {\bibinfo
  {journal} {Nature Nanotechnology}\ }\textbf {\bibinfo {volume} {13}},\
  \bibinfo {pages} {53} (\bibinfo {year} {2018})}\BibitemShut {NoStop}%
\bibitem [{\citenamefont {Mengotti}\ \emph {et~al.}(2011)\citenamefont
  {Mengotti}, \citenamefont {Heyderman}, \citenamefont {Rodriguez},
  \citenamefont {Nolting}, \citenamefont {Hugli},\ and\ \citenamefont
  {Braun}}]{2010Mengotti}%
  \BibitemOpen
  \bibfield  {author} {\bibinfo {author} {\bibfnamefont {E.}~\bibnamefont
  {Mengotti}}, \bibinfo {author} {\bibfnamefont {L.~J.}\ \bibnamefont
  {Heyderman}}, \bibinfo {author} {\bibfnamefont {A.~F.}\ \bibnamefont
  {Rodriguez}}, \bibinfo {author} {\bibfnamefont {F.}~\bibnamefont {Nolting}},
  \bibinfo {author} {\bibfnamefont {R.~V.}\ \bibnamefont {Hugli}},\ and\
  \bibinfo {author} {\bibfnamefont {H.-B.}\ \bibnamefont {Braun}},\ }\bibfield
  {title} {\bibinfo {title} {Real-space observation of emergent magnetic
  monopoles and associated {Dirac} strings in artificial kagome spin ice},\
  }\href@noop {} {\bibfield  {journal} {\bibinfo  {journal} {Nat Phys}\
  }\textbf {\bibinfo {volume} {7}},\ \bibinfo {pages} {68} (\bibinfo {year}
  {2011})}\BibitemShut {NoStop}%
\bibitem [{\citenamefont {Rougemaille}\ \emph {et~al.}(2013)\citenamefont
  {Rougemaille}, \citenamefont {Montaigne}, \citenamefont {Canals},
  \citenamefont {Hehn}, \citenamefont {Riahi}, \citenamefont {Lacour},\ and\
  \citenamefont {Toussaint}}]{2013Rougemaille}%
  \BibitemOpen
  \bibfield  {author} {\bibinfo {author} {\bibfnamefont {N.}~\bibnamefont
  {Rougemaille}}, \bibinfo {author} {\bibfnamefont {F.}~\bibnamefont
  {Montaigne}}, \bibinfo {author} {\bibfnamefont {B.}~\bibnamefont {Canals}},
  \bibinfo {author} {\bibfnamefont {M.}~\bibnamefont {Hehn}}, \bibinfo {author}
  {\bibfnamefont {H.}~\bibnamefont {Riahi}}, \bibinfo {author} {\bibfnamefont
  {D.}~\bibnamefont {Lacour}},\ and\ \bibinfo {author} {\bibfnamefont {J.-C.}\
  \bibnamefont {Toussaint}},\ }\bibfield  {title} {\bibinfo {title} {Chiral
  nature of magnetic monopoles in artificial spin ice},\ }\href
  {https://doi.org/10.1088/1367-2630/15/3/035026} {\bibfield  {journal}
  {\bibinfo  {journal} {New Journal of Physics}\ }\textbf {\bibinfo {volume}
  {15}},\ \bibinfo {pages} {035026} (\bibinfo {year} {2013})}\BibitemShut
  {NoStop}%
\bibitem [{\citenamefont {Le}\ \emph {et~al.}(2017)\citenamefont {Le},
  \citenamefont {Park}, \citenamefont {Sklenar}, \citenamefont {Chern},
  \citenamefont {Nisoli}, \citenamefont {Watts}, \citenamefont {Manno},
  \citenamefont {Rench}, \citenamefont {Samarth}, \citenamefont {Leighton},\
  and\ \citenamefont {Schiffer}}]{2017Le}%
  \BibitemOpen
  \bibfield  {author} {\bibinfo {author} {\bibfnamefont {B.~L.}\ \bibnamefont
  {Le}}, \bibinfo {author} {\bibfnamefont {J.}~\bibnamefont {Park}}, \bibinfo
  {author} {\bibfnamefont {J.}~\bibnamefont {Sklenar}}, \bibinfo {author}
  {\bibfnamefont {G.-W.}\ \bibnamefont {Chern}}, \bibinfo {author}
  {\bibfnamefont {C.}~\bibnamefont {Nisoli}}, \bibinfo {author} {\bibfnamefont
  {J.~D.}\ \bibnamefont {Watts}}, \bibinfo {author} {\bibfnamefont
  {M.}~\bibnamefont {Manno}}, \bibinfo {author} {\bibfnamefont {D.~W.}\
  \bibnamefont {Rench}}, \bibinfo {author} {\bibfnamefont {N.}~\bibnamefont
  {Samarth}}, \bibinfo {author} {\bibfnamefont {C.}~\bibnamefont {Leighton}},\
  and\ \bibinfo {author} {\bibfnamefont {P.}~\bibnamefont {Schiffer}},\
  }\bibfield  {title} {\bibinfo {title} {Understanding magnetotransport
  signatures in networks of connected permalloy nanowires},\ }\href
  {https://doi.org/10.1103/PhysRevB.95.060405} {\bibfield  {journal} {\bibinfo
  {journal} {Phys. Rev. B}\ }\textbf {\bibinfo {volume} {95}},\ \bibinfo
  {pages} {060405} (\bibinfo {year} {2017})}\BibitemShut {NoStop}%
\bibitem [{\citenamefont {Dawidek}\ \emph {et~al.}(2021)\citenamefont
  {Dawidek}, \citenamefont {Hayward}, \citenamefont {Vidamour}, \citenamefont
  {Broomhall}, \citenamefont {Venkat}, \citenamefont {Mamoori}, \citenamefont
  {Mullen}, \citenamefont {Kyle}, \citenamefont {Fry}, \citenamefont {Steinke},
  \citenamefont {Cooper}, \citenamefont {Maccherozzi}, \citenamefont {Dhesi},
  \citenamefont {Aballe}, \citenamefont {Foerster}, \citenamefont {Prat},
  \citenamefont {Vasilaki}, \citenamefont {Ellis},\ and\ \citenamefont
  {Allwood}}]{2021Dawidek}%
  \BibitemOpen
  \bibfield  {author} {\bibinfo {author} {\bibfnamefont {R.~W.}\ \bibnamefont
  {Dawidek}}, \bibinfo {author} {\bibfnamefont {T.~J.}\ \bibnamefont
  {Hayward}}, \bibinfo {author} {\bibfnamefont {I.~T.}\ \bibnamefont
  {Vidamour}}, \bibinfo {author} {\bibfnamefont {T.~J.}\ \bibnamefont
  {Broomhall}}, \bibinfo {author} {\bibfnamefont {G.}~\bibnamefont {Venkat}},
  \bibinfo {author} {\bibfnamefont {M.~A.}\ \bibnamefont {Mamoori}}, \bibinfo
  {author} {\bibfnamefont {A.}~\bibnamefont {Mullen}}, \bibinfo {author}
  {\bibfnamefont {S.~J.}\ \bibnamefont {Kyle}}, \bibinfo {author}
  {\bibfnamefont {P.~W.}\ \bibnamefont {Fry}}, \bibinfo {author} {\bibfnamefont
  {N.-J.}\ \bibnamefont {Steinke}}, \bibinfo {author} {\bibfnamefont
  {J.~F.~K.}\ \bibnamefont {Cooper}}, \bibinfo {author} {\bibfnamefont
  {F.}~\bibnamefont {Maccherozzi}}, \bibinfo {author} {\bibfnamefont {S.~S.}\
  \bibnamefont {Dhesi}}, \bibinfo {author} {\bibfnamefont {L.}~\bibnamefont
  {Aballe}}, \bibinfo {author} {\bibfnamefont {M.}~\bibnamefont {Foerster}},
  \bibinfo {author} {\bibfnamefont {J.}~\bibnamefont {Prat}}, \bibinfo {author}
  {\bibfnamefont {E.}~\bibnamefont {Vasilaki}}, \bibinfo {author}
  {\bibfnamefont {M.~O.~A.}\ \bibnamefont {Ellis}},\ and\ \bibinfo {author}
  {\bibfnamefont {D.~A.}\ \bibnamefont {Allwood}},\ }\bibfield  {title}
  {\bibinfo {title} {Dynamically-driven emergence in a nanomagnetic system},\
  }\href {https://doi.org/https://doi.org/10.1002/adfm.202008389} {\bibfield
  {journal} {\bibinfo  {journal} {Advanced Functional Materials}\ }\textbf
  {\bibinfo {volume} {31}},\ \bibinfo {pages} {2008389} (\bibinfo {year}
  {2021})}\BibitemShut {NoStop}%
\bibitem [{\citenamefont {Gartside}\ \emph {et~al.}(2022)\citenamefont
  {Gartside}, \citenamefont {Stenning}, \citenamefont {Vanstone}, \citenamefont
  {Holder}, \citenamefont {Arroo}, \citenamefont {Dion}, \citenamefont
  {Caravelli}, \citenamefont {Kurebayashi},\ and\ \citenamefont
  {Branford}}]{2022Gartside}%
  \BibitemOpen
  \bibfield  {author} {\bibinfo {author} {\bibfnamefont {J.~C.}\ \bibnamefont
  {Gartside}}, \bibinfo {author} {\bibfnamefont {K.~D.}\ \bibnamefont
  {Stenning}}, \bibinfo {author} {\bibfnamefont {A.}~\bibnamefont {Vanstone}},
  \bibinfo {author} {\bibfnamefont {H.~H.}\ \bibnamefont {Holder}}, \bibinfo
  {author} {\bibfnamefont {D.~M.}\ \bibnamefont {Arroo}}, \bibinfo {author}
  {\bibfnamefont {T.}~\bibnamefont {Dion}}, \bibinfo {author} {\bibfnamefont
  {F.}~\bibnamefont {Caravelli}}, \bibinfo {author} {\bibfnamefont
  {H.}~\bibnamefont {Kurebayashi}},\ and\ \bibinfo {author} {\bibfnamefont
  {W.~R.}\ \bibnamefont {Branford}},\ }\bibfield  {title} {\bibinfo {title}
  {Reconfigurable training and reservoir computing in an artificial spin-vortex
  ice via spin-wave fingerprinting},\ }\href
  {https://doi.org/10.1038/s41565-022-01091-7} {\bibfield  {journal} {\bibinfo
  {journal} {Nature Nanotechnology}\ }\textbf {\bibinfo {volume} {17}},\
  \bibinfo {pages} {460} (\bibinfo {year} {2022})}\BibitemShut {NoStop}%
\bibitem [{\citenamefont {Bhattacharya}\ \emph {et~al.}(2 12)\citenamefont
  {Bhattacharya}, \citenamefont {Chen}, \citenamefont {Jensen}, \citenamefont
  {Liu}, \citenamefont {Burks}, \citenamefont {Gilbert}, \citenamefont {Zhang},
  \citenamefont {Yin},\ and\ \citenamefont {Liu}}]{2022Bhattacharya}%
  \BibitemOpen
  \bibfield  {author} {\bibinfo {author} {\bibfnamefont {D.}~\bibnamefont
  {Bhattacharya}}, \bibinfo {author} {\bibfnamefont {Z.}~\bibnamefont {Chen}},
  \bibinfo {author} {\bibfnamefont {C.~J.}\ \bibnamefont {Jensen}}, \bibinfo
  {author} {\bibfnamefont {C.}~\bibnamefont {Liu}}, \bibinfo {author}
  {\bibfnamefont {E.~C.}\ \bibnamefont {Burks}}, \bibinfo {author}
  {\bibfnamefont {D.~A.}\ \bibnamefont {Gilbert}}, \bibinfo {author}
  {\bibfnamefont {X.}~\bibnamefont {Zhang}}, \bibinfo {author} {\bibfnamefont
  {G.}~\bibnamefont {Yin}},\ and\ \bibinfo {author} {\bibfnamefont
  {K.}~\bibnamefont {Liu}},\ }\bibfield  {title} {\bibinfo {title} {3d
  interconnected magnetic nanowire networks as potential integrated multistate
  memristors},\ }\href {https://doi.org/10.1021/acs.nanolett.2c03616}
  {\bibfield  {journal} {\bibinfo  {journal} {Nano Lett.}\ }\textbf {\bibinfo
  {volume} {22}},\ \bibinfo {pages} {10010} (\bibinfo {year}
  {2022-12})}\BibitemShut {NoStop}%
\bibitem [{\citenamefont {Chern}\ \emph {et~al.}(2014)\citenamefont {Chern},
  \citenamefont {Reichhardt},\ and\ \citenamefont {Nisoli}}]{2014Chern_a}%
  \BibitemOpen
  \bibfield  {author} {\bibinfo {author} {\bibfnamefont {G.-W.}\ \bibnamefont
  {Chern}}, \bibinfo {author} {\bibfnamefont {C.}~\bibnamefont {Reichhardt}},\
  and\ \bibinfo {author} {\bibfnamefont {C.}~\bibnamefont {Nisoli}},\
  }\bibfield  {title} {\bibinfo {title} {Realizing three-dimensional artificial
  spin ice by stacking planar nano-arrays},\ }\href
  {https://doi.org/http://dx.doi.org/10.1063/1.4861118} {\bibfield  {journal}
  {\bibinfo  {journal} {Applied Physics Letters}\ }\textbf {\bibinfo {volume}
  {104}},\ \bibinfo {eid} {013101} (\bibinfo {year} {2014})}\BibitemShut
  {NoStop}%
\bibitem [{\citenamefont {Perrin}\ \emph {et~al.}(2016)\citenamefont {Perrin},
  \citenamefont {Canals},\ and\ \citenamefont {Rougemaille}}]{2016Perrin}%
  \BibitemOpen
  \bibfield  {author} {\bibinfo {author} {\bibfnamefont {Y.}~\bibnamefont
  {Perrin}}, \bibinfo {author} {\bibfnamefont {B.}~\bibnamefont {Canals}},\
  and\ \bibinfo {author} {\bibfnamefont {N.}~\bibnamefont {Rougemaille}},\
  }\bibfield  {title} {\bibinfo {title} {Extensive degeneracy, coulomb phase
  and magnetic monopoles in artificial square ice},\ }\href
  {http://dx.doi.org/10.1038/nature20155} {\bibfield  {journal} {\bibinfo
  {journal} {Nature}\ }\textbf {\bibinfo {volume} {540}},\ \bibinfo {pages}
  {410} (\bibinfo {year} {2016})}\BibitemShut {NoStop}%
\bibitem [{\citenamefont {Fernández-Pacheco}\ \emph
  {et~al.}(2017)\citenamefont {Fernández-Pacheco}, \citenamefont {Streubel},
  \citenamefont {Fruchart}, \citenamefont {Hertel}, \citenamefont {Fischer},\
  and\ \citenamefont {Cowburn}}]{2017Fernandez}%
  \BibitemOpen
  \bibfield  {author} {\bibinfo {author} {\bibfnamefont {A.}~\bibnamefont
  {Fernández-Pacheco}}, \bibinfo {author} {\bibfnamefont {R.}~\bibnamefont
  {Streubel}}, \bibinfo {author} {\bibfnamefont {O.}~\bibnamefont {Fruchart}},
  \bibinfo {author} {\bibfnamefont {R.}~\bibnamefont {Hertel}}, \bibinfo
  {author} {\bibfnamefont {P.}~\bibnamefont {Fischer}},\ and\ \bibinfo {author}
  {\bibfnamefont {R.~P.}\ \bibnamefont {Cowburn}},\ }\bibfield  {title}
  {\bibinfo {title} {Three-dimensional nanomagnetism},\ }\href
  {https://doi.org/10.1038/ncomms15756} {\bibfield  {journal} {\bibinfo
  {journal} {Nature Communications}\ }\textbf {\bibinfo {volume} {8}},\
  \bibinfo {pages} {15756} (\bibinfo {year} {2017})}\BibitemShut {NoStop}%
\bibitem [{\citenamefont {Makarov}\ \emph {et~al.}(2022)\citenamefont
  {Makarov}, \citenamefont {Volkov}, \citenamefont {K{\'{a}}kay}, \citenamefont
  {Pylypovskyi}, \citenamefont {Budinsk{\'{a}}},\ and\ \citenamefont
  {Dobrovolskiy}}]{2022Makarov_a}%
  \BibitemOpen
  \bibfield  {author} {\bibinfo {author} {\bibfnamefont {D.}~\bibnamefont
  {Makarov}}, \bibinfo {author} {\bibfnamefont {O.~M.}\ \bibnamefont {Volkov}},
  \bibinfo {author} {\bibfnamefont {A.}~\bibnamefont {K{\'{a}}kay}}, \bibinfo
  {author} {\bibfnamefont {O.~V.}\ \bibnamefont {Pylypovskyi}}, \bibinfo
  {author} {\bibfnamefont {B.}~\bibnamefont {Budinsk{\'{a}}}},\ and\ \bibinfo
  {author} {\bibfnamefont {O.~V.}\ \bibnamefont {Dobrovolskiy}},\ }\bibfield
  {title} {\bibinfo {title} {{New Dimension in Magnetism and Superconductivity:
  3D and Curvilinear Nanoarchitectures}},\ }\href
  {https://doi.org/10.1002/adma.202101758} {\bibfield  {journal} {\bibinfo
  {journal} {Advanced Materials}\ }\textbf {\bibinfo {volume} {34}},\ \bibinfo
  {pages} {2101758} (\bibinfo {year} {2022})}\BibitemShut {NoStop}%
\bibitem [{\citenamefont {Streubel}\ \emph {et~al.}(2016)\citenamefont
  {Streubel}, \citenamefont {Fischer}, \citenamefont {Kronast}, \citenamefont
  {Kravchuk}, \citenamefont {Sheka}, \citenamefont {Gaididei}, \citenamefont
  {Schmidt},\ and\ \citenamefont {Makarov}}]{2016Streubel}%
  \BibitemOpen
  \bibfield  {author} {\bibinfo {author} {\bibfnamefont {R.}~\bibnamefont
  {Streubel}}, \bibinfo {author} {\bibfnamefont {P.}~\bibnamefont {Fischer}},
  \bibinfo {author} {\bibfnamefont {F.}~\bibnamefont {Kronast}}, \bibinfo
  {author} {\bibfnamefont {V.~P.}\ \bibnamefont {Kravchuk}}, \bibinfo {author}
  {\bibfnamefont {D.~D.}\ \bibnamefont {Sheka}}, \bibinfo {author}
  {\bibfnamefont {Y.}~\bibnamefont {Gaididei}}, \bibinfo {author}
  {\bibfnamefont {O.~G.}\ \bibnamefont {Schmidt}},\ and\ \bibinfo {author}
  {\bibfnamefont {D.}~\bibnamefont {Makarov}},\ }\bibfield  {title} {\bibinfo
  {title} {Magnetism in curved geometries},\ }\href
  {https://doi.org/10.1088/0022-3727/49/36/363001} {\bibfield  {journal}
  {\bibinfo  {journal} {Journal of Physics D: Applied Physics}\ }\textbf
  {\bibinfo {volume} {49}},\ \bibinfo {pages} {363001} (\bibinfo {year}
  {2016})}\BibitemShut {NoStop}%
\bibitem [{\citenamefont {Volkov}\ \emph {et~al.}(2018)\citenamefont {Volkov},
  \citenamefont {Sheka}, \citenamefont {Gaididei}, \citenamefont {Kravchuk},
  \citenamefont {R{\"{o}}{\ss}ler}, \citenamefont {Fassbender},\ and\
  \citenamefont {Makarov}}]{2018Volkov}%
  \BibitemOpen
  \bibfield  {author} {\bibinfo {author} {\bibfnamefont {O.~M.}\ \bibnamefont
  {Volkov}}, \bibinfo {author} {\bibfnamefont {D.~D.}\ \bibnamefont {Sheka}},
  \bibinfo {author} {\bibfnamefont {Y.}~\bibnamefont {Gaididei}}, \bibinfo
  {author} {\bibfnamefont {V.~P.}\ \bibnamefont {Kravchuk}}, \bibinfo {author}
  {\bibfnamefont {U.~K.}\ \bibnamefont {R{\"{o}}{\ss}ler}}, \bibinfo {author}
  {\bibfnamefont {J.}~\bibnamefont {Fassbender}},\ and\ \bibinfo {author}
  {\bibfnamefont {D.}~\bibnamefont {Makarov}},\ }\bibfield  {title} {\bibinfo
  {title} {Mesoscale {Dzyaloshinskii-Moriya} interaction: {G}eometrical
  tailoring of the magnetochirality},\ }\href
  {https://doi.org/10.1038/s41598-017-18835-4} {\bibfield  {journal} {\bibinfo
  {journal} {Scientific Reports}\ }\textbf {\bibinfo {volume} {8}},\ \bibinfo
  {pages} {866} (\bibinfo {year} {2018})}\BibitemShut {NoStop}%
\bibitem [{\citenamefont {Sheka}\ \emph {et~al.}(2020)\citenamefont {Sheka},
  \citenamefont {Pylypovskyi}, \citenamefont {Landeros}, \citenamefont
  {Gaididei}, \citenamefont {Kákay},\ and\ \citenamefont
  {Makarov}}]{2020Sheka}%
  \BibitemOpen
  \bibfield  {author} {\bibinfo {author} {\bibfnamefont {D.~D.}\ \bibnamefont
  {Sheka}}, \bibinfo {author} {\bibfnamefont {O.~V.}\ \bibnamefont
  {Pylypovskyi}}, \bibinfo {author} {\bibfnamefont {P.}~\bibnamefont
  {Landeros}}, \bibinfo {author} {\bibfnamefont {Y.}~\bibnamefont {Gaididei}},
  \bibinfo {author} {\bibfnamefont {A.}~\bibnamefont {Kákay}},\ and\ \bibinfo
  {author} {\bibfnamefont {D.}~\bibnamefont {Makarov}},\ }\bibfield  {title}
  {\bibinfo {title} {Nonlocal chiral symmetry breaking in curvilinear magnetic
  shells},\ }\href {https://doi.org/10.1038/s42005-020-0387-2} {\bibfield
  {journal} {\bibinfo  {journal} {Communications Physics}\ }\textbf {\bibinfo
  {volume} {3}},\ \bibinfo {pages} {128} (\bibinfo {year} {2020})}\BibitemShut
  {NoStop}%
\bibitem [{\citenamefont {Sahoo}\ \emph {et~al.}(2021)\citenamefont {Sahoo},
  \citenamefont {May}, \citenamefont {van Den~Berg}, \citenamefont {Mondal},
  \citenamefont {Ladak},\ and\ \citenamefont {Barman}}]{2021Sahoo}%
  \BibitemOpen
  \bibfield  {author} {\bibinfo {author} {\bibfnamefont {S.}~\bibnamefont
  {Sahoo}}, \bibinfo {author} {\bibfnamefont {A.}~\bibnamefont {May}}, \bibinfo
  {author} {\bibfnamefont {A.}~\bibnamefont {van Den~Berg}}, \bibinfo {author}
  {\bibfnamefont {A.~K.}\ \bibnamefont {Mondal}}, \bibinfo {author}
  {\bibfnamefont {S.}~\bibnamefont {Ladak}},\ and\ \bibinfo {author}
  {\bibfnamefont {A.}~\bibnamefont {Barman}},\ }\bibfield  {title} {\bibinfo
  {title} {Observation of coherent spin waves in a three-dimensional artificial
  spin ice structure},\ }\href {https://doi.org/10.1021/acs.nanolett.1c00650}
  {\bibfield  {journal} {\bibinfo  {journal} {Nano Lett.}\ }\textbf {\bibinfo
  {volume} {21}},\ \bibinfo {pages} {4629} (\bibinfo {year}
  {2021})}\BibitemShut {NoStop}%
\bibitem [{\citenamefont {Cheenikundil}\ \emph {et~al.}(2022)\citenamefont
  {Cheenikundil}, \citenamefont {Bauer}, \citenamefont {Goharyan},
  \citenamefont {D'Aquino},\ and\ \citenamefont {Hertel}}]{2022Cheenikundil}%
  \BibitemOpen
  \bibfield  {author} {\bibinfo {author} {\bibfnamefont {R.}~\bibnamefont
  {Cheenikundil}}, \bibinfo {author} {\bibfnamefont {J.}~\bibnamefont {Bauer}},
  \bibinfo {author} {\bibfnamefont {M.}~\bibnamefont {Goharyan}}, \bibinfo
  {author} {\bibfnamefont {M.}~\bibnamefont {D'Aquino}},\ and\ \bibinfo
  {author} {\bibfnamefont {R.}~\bibnamefont {Hertel}},\ }\bibfield  {title}
  {\bibinfo {title} {{High-frequency modes in a magnetic buckyball
  nanoarchitecture}},\ }\href {https://doi.org/10.1063/5.0097695} {\bibfield
  {journal} {\bibinfo  {journal} {APL Materials}\ }\textbf {\bibinfo {volume}
  {10}},\ \bibinfo {pages} {081106} (\bibinfo {year} {2022})}\BibitemShut
  {NoStop}%
\bibitem [{\citenamefont {K\"orber}\ \emph {et~al.}(2022)\citenamefont
  {K\"orber}, \citenamefont {Verba}, \citenamefont {Ot\'alora}, \citenamefont
  {Kravchuk}, \citenamefont {Lindner}, \citenamefont {Fassbender},\ and\
  \citenamefont {K\'akay}}]{2022Koerber}%
  \BibitemOpen
  \bibfield  {author} {\bibinfo {author} {\bibfnamefont {L.}~\bibnamefont
  {K\"orber}}, \bibinfo {author} {\bibfnamefont {R.}~\bibnamefont {Verba}},
  \bibinfo {author} {\bibfnamefont {J.~A.}\ \bibnamefont {Ot\'alora}}, \bibinfo
  {author} {\bibfnamefont {V.}~\bibnamefont {Kravchuk}}, \bibinfo {author}
  {\bibfnamefont {J.}~\bibnamefont {Lindner}}, \bibinfo {author} {\bibfnamefont
  {J.}~\bibnamefont {Fassbender}},\ and\ \bibinfo {author} {\bibfnamefont
  {A.}~\bibnamefont {K\'akay}},\ }\bibfield  {title} {\bibinfo {title}
  {Curvilinear spin-wave dynamics beyond the thin-shell approximation: Magnetic
  nanotubes as a case study},\ }\href
  {https://doi.org/10.1103/PhysRevB.106.014405} {\bibfield  {journal} {\bibinfo
   {journal} {Phys. Rev. B}\ }\textbf {\bibinfo {volume} {106}},\ \bibinfo
  {pages} {014405} (\bibinfo {year} {2022})}\BibitemShut {NoStop}%
\bibitem [{\citenamefont {Skoric}\ \emph {et~al.}(2022)\citenamefont {Skoric},
  \citenamefont {Donnelly}, \citenamefont {Hierro-Rodriguez}, \citenamefont
  {Cascales~Sandoval}, \citenamefont {Ruiz-Gómez}, \citenamefont {Foerster},
  \citenamefont {Niño}, \citenamefont {Belkhou}, \citenamefont {Abert},
  \citenamefont {Suess},\ and\ \citenamefont
  {Fernández-Pacheco}}]{2022Skoric}%
  \BibitemOpen
  \bibfield  {author} {\bibinfo {author} {\bibfnamefont {L.}~\bibnamefont
  {Skoric}}, \bibinfo {author} {\bibfnamefont {C.}~\bibnamefont {Donnelly}},
  \bibinfo {author} {\bibfnamefont {A.}~\bibnamefont {Hierro-Rodriguez}},
  \bibinfo {author} {\bibfnamefont {M.~A.}\ \bibnamefont {Cascales~Sandoval}},
  \bibinfo {author} {\bibfnamefont {S.}~\bibnamefont {Ruiz-Gómez}}, \bibinfo
  {author} {\bibfnamefont {M.}~\bibnamefont {Foerster}}, \bibinfo {author}
  {\bibfnamefont {M.~A.}\ \bibnamefont {Niño}}, \bibinfo {author}
  {\bibfnamefont {R.}~\bibnamefont {Belkhou}}, \bibinfo {author} {\bibfnamefont
  {C.}~\bibnamefont {Abert}}, \bibinfo {author} {\bibfnamefont
  {D.}~\bibnamefont {Suess}},\ and\ \bibinfo {author} {\bibfnamefont
  {A.}~\bibnamefont {Fernández-Pacheco}},\ }\bibfield  {title} {\bibinfo
  {title} {Domain wall automotion in three-dimensional magnetic helical
  interconnectors},\ }\href {https://doi.org/10.1021/acsnano.1c10345}
  {\bibfield  {journal} {\bibinfo  {journal} {ACS Nano}\ }\textbf {\bibinfo
  {volume} {16}},\ \bibinfo {pages} {8860} (\bibinfo {year}
  {2022})}\BibitemShut {NoStop}%
\bibitem [{\citenamefont {May}\ \emph {et~al.}(2021)\citenamefont {May},
  \citenamefont {Saccone}, \citenamefont {van~den Berg}, \citenamefont {Askey},
  \citenamefont {Hunt},\ and\ \citenamefont {Ladak}}]{2021May}%
  \BibitemOpen
  \bibfield  {author} {\bibinfo {author} {\bibfnamefont {A.}~\bibnamefont
  {May}}, \bibinfo {author} {\bibfnamefont {M.}~\bibnamefont {Saccone}},
  \bibinfo {author} {\bibfnamefont {A.}~\bibnamefont {van~den Berg}}, \bibinfo
  {author} {\bibfnamefont {J.}~\bibnamefont {Askey}}, \bibinfo {author}
  {\bibfnamefont {M.}~\bibnamefont {Hunt}},\ and\ \bibinfo {author}
  {\bibfnamefont {S.}~\bibnamefont {Ladak}},\ }\bibfield  {title} {\bibinfo
  {title} {Magnetic charge propagation upon a {3D} artificial spin-ice},\
  }\href {https://doi.org/10.1038/s41467-021-23480-7} {\bibfield  {journal}
  {\bibinfo  {journal} {Nature Communications}\ }\textbf {\bibinfo {volume}
  {12}},\ \bibinfo {pages} {3217} (\bibinfo {year} {2021})}\BibitemShut
  {NoStop}%
\bibitem [{\citenamefont {Koraltan}\ \emph {et~al.}(2021)\citenamefont
  {Koraltan}, \citenamefont {Slanovc}, \citenamefont {Bruckner}, \citenamefont
  {Nisoli}, \citenamefont {Chumak}, \citenamefont {Dobrovolskiy}, \citenamefont
  {Abert},\ and\ \citenamefont {Suess}}]{2021Koraltan}%
  \BibitemOpen
  \bibfield  {author} {\bibinfo {author} {\bibfnamefont {S.}~\bibnamefont
  {Koraltan}}, \bibinfo {author} {\bibfnamefont {F.}~\bibnamefont {Slanovc}},
  \bibinfo {author} {\bibfnamefont {F.}~\bibnamefont {Bruckner}}, \bibinfo
  {author} {\bibfnamefont {C.}~\bibnamefont {Nisoli}}, \bibinfo {author}
  {\bibfnamefont {A.~V.}\ \bibnamefont {Chumak}}, \bibinfo {author}
  {\bibfnamefont {O.~V.}\ \bibnamefont {Dobrovolskiy}}, \bibinfo {author}
  {\bibfnamefont {C.}~\bibnamefont {Abert}},\ and\ \bibinfo {author}
  {\bibfnamefont {D.}~\bibnamefont {Suess}},\ }\bibfield  {title} {\bibinfo
  {title} {Tension-free {Dirac} strings and steered magnetic charges in {3D}
  artificial spin ice},\ }\href {https://doi.org/10.1038/s41524-021-00593-7}
  {\bibfield  {journal} {\bibinfo  {journal} {npj Computational Materials}\
  }\textbf {\bibinfo {volume} {7}},\ \bibinfo {pages} {125} (\bibinfo {year}
  {2021})}\BibitemShut {NoStop}%
\bibitem [{\citenamefont {Pip}\ \emph {et~al.}(2022)\citenamefont {Pip},
  \citenamefont {Treves}, \citenamefont {Massey}, \citenamefont {Finizio},
  \citenamefont {Luo}, \citenamefont {Hrabec}, \citenamefont {Scagnoli},
  \citenamefont {Raabe}, \citenamefont {Philippe}, \citenamefont {Heyderman},\
  and\ \citenamefont {Donnelly}}]{2022Pip}%
  \BibitemOpen
  \bibfield  {author} {\bibinfo {author} {\bibfnamefont {P.}~\bibnamefont
  {Pip}}, \bibinfo {author} {\bibfnamefont {S.}~\bibnamefont {Treves}},
  \bibinfo {author} {\bibfnamefont {J.~R.}\ \bibnamefont {Massey}}, \bibinfo
  {author} {\bibfnamefont {S.}~\bibnamefont {Finizio}}, \bibinfo {author}
  {\bibfnamefont {Z.}~\bibnamefont {Luo}}, \bibinfo {author} {\bibfnamefont
  {A.}~\bibnamefont {Hrabec}}, \bibinfo {author} {\bibfnamefont
  {V.}~\bibnamefont {Scagnoli}}, \bibinfo {author} {\bibfnamefont
  {J.}~\bibnamefont {Raabe}}, \bibinfo {author} {\bibfnamefont
  {L.}~\bibnamefont {Philippe}}, \bibinfo {author} {\bibfnamefont {L.~J.}\
  \bibnamefont {Heyderman}},\ and\ \bibinfo {author} {\bibfnamefont
  {C.}~\bibnamefont {Donnelly}},\ }\bibfield  {title} {\bibinfo {title} {X-ray
  imaging of the magnetic configuration of a three-dimensional artificial spin
  ice building block},\ }\href {https://doi.org/10.1063/5.0101797} {\bibfield
  {journal} {\bibinfo  {journal} {APL Materials}\ }\textbf {\bibinfo {volume}
  {10}},\ \bibinfo {pages} {101101} (\bibinfo {year} {2022})}\BibitemShut
  {NoStop}%
\bibitem [{\citenamefont {Grigoryeva}\ \emph {et~al.}(2011)\citenamefont
  {Grigoryeva}, \citenamefont {Mistonov}, \citenamefont {Napolskii},
  \citenamefont {Sapoletova}, \citenamefont {Eliseev}, \citenamefont {Bouwman},
  \citenamefont {Byelov}, \citenamefont {Petukhov}, \citenamefont {Chernyshov},
  \citenamefont {Eckerlebe}, \citenamefont {Vasilieva},\ and\ \citenamefont
  {Grigoriev}}]{2011Grigoryeva}%
  \BibitemOpen
  \bibfield  {author} {\bibinfo {author} {\bibfnamefont {N.~A.}\ \bibnamefont
  {Grigoryeva}}, \bibinfo {author} {\bibfnamefont {A.~A.}\ \bibnamefont
  {Mistonov}}, \bibinfo {author} {\bibfnamefont {K.~S.}\ \bibnamefont
  {Napolskii}}, \bibinfo {author} {\bibfnamefont {N.~A.}\ \bibnamefont
  {Sapoletova}}, \bibinfo {author} {\bibfnamefont {A.~A.}\ \bibnamefont
  {Eliseev}}, \bibinfo {author} {\bibfnamefont {W.}~\bibnamefont {Bouwman}},
  \bibinfo {author} {\bibfnamefont {D.~V.}\ \bibnamefont {Byelov}}, \bibinfo
  {author} {\bibfnamefont {A.~V.}\ \bibnamefont {Petukhov}}, \bibinfo {author}
  {\bibfnamefont {D.~Y.}\ \bibnamefont {Chernyshov}}, \bibinfo {author}
  {\bibfnamefont {H.}~\bibnamefont {Eckerlebe}}, \bibinfo {author}
  {\bibfnamefont {A.~V.}\ \bibnamefont {Vasilieva}},\ and\ \bibinfo {author}
  {\bibfnamefont {S.~V.}\ \bibnamefont {Grigoriev}},\ }\bibfield  {title}
  {\bibinfo {title} {Magnetic topology of {Co}-based inverse opal-like
  structures},\ }\href {https://doi.org/10.1103/PhysRevB.84.064405} {\bibfield
  {journal} {\bibinfo  {journal} {Phys. Rev. B}\ }\textbf {\bibinfo {volume}
  {84}},\ \bibinfo {pages} {064405} (\bibinfo {year} {2011})}\BibitemShut
  {NoStop}%
\bibitem [{\citenamefont {Shishkin}\ \emph {et~al.}(2016)\citenamefont
  {Shishkin}, \citenamefont {Mistonov}, \citenamefont {Dubitskiy},
  \citenamefont {Grigoryeva}, \citenamefont {Menzel},\ and\ \citenamefont
  {Grigoriev}}]{2016Shishkin}%
  \BibitemOpen
  \bibfield  {author} {\bibinfo {author} {\bibfnamefont {I.~S.}\ \bibnamefont
  {Shishkin}}, \bibinfo {author} {\bibfnamefont {A.~A.}\ \bibnamefont
  {Mistonov}}, \bibinfo {author} {\bibfnamefont {I.~S.}\ \bibnamefont
  {Dubitskiy}}, \bibinfo {author} {\bibfnamefont {N.~A.}\ \bibnamefont
  {Grigoryeva}}, \bibinfo {author} {\bibfnamefont {D.}~\bibnamefont {Menzel}},\
  and\ \bibinfo {author} {\bibfnamefont {S.~V.}\ \bibnamefont {Grigoriev}},\
  }\bibfield  {title} {\bibinfo {title} {Nonlinear geometric scaling of
  coercivity in a three-dimensional nanoscale analog of spin ice},\ }\href
  {https://doi.org/10.1103/PhysRevB.94.064424} {\bibfield  {journal} {\bibinfo
  {journal} {Phys. Rev. B}\ }\textbf {\bibinfo {volume} {94}},\ \bibinfo
  {pages} {064424} (\bibinfo {year} {2016})}\BibitemShut {NoStop}%
\bibitem [{\citenamefont {Mistonov}\ \emph {et~al.}(2019)\citenamefont
  {Mistonov}, \citenamefont {Dubitskiy}, \citenamefont {Shishkin},
  \citenamefont {Grigoryeva}, \citenamefont {Heinemann}, \citenamefont
  {Sapoletova}, \citenamefont {Valkovskiy},\ and\ \citenamefont
  {Grigoriev}}]{2019Mistonov}%
  \BibitemOpen
  \bibfield  {author} {\bibinfo {author} {\bibfnamefont {A.~A.}\ \bibnamefont
  {Mistonov}}, \bibinfo {author} {\bibfnamefont {I.~S.}\ \bibnamefont
  {Dubitskiy}}, \bibinfo {author} {\bibfnamefont {I.~S.}\ \bibnamefont
  {Shishkin}}, \bibinfo {author} {\bibfnamefont {N.~A.}\ \bibnamefont
  {Grigoryeva}}, \bibinfo {author} {\bibfnamefont {A.}~\bibnamefont
  {Heinemann}}, \bibinfo {author} {\bibfnamefont {N.~A.}\ \bibnamefont
  {Sapoletova}}, \bibinfo {author} {\bibfnamefont {G.~A.}\ \bibnamefont
  {Valkovskiy}},\ and\ \bibinfo {author} {\bibfnamefont {S.~V.}\ \bibnamefont
  {Grigoriev}},\ }\bibfield  {title} {\bibinfo {title} {Magnetic structure of
  the inverse opal-like structures: Small angle neutron diffraction and
  micromagnetic simulations},\ }\href
  {https://doi.org/10.1016/j.jmmm.2019.01.016} {\bibfield  {journal} {\bibinfo
  {journal} {Journal of Magnetism and Magnetic Materials}\ }\textbf {\bibinfo
  {volume} {477}},\ \bibinfo {pages} {99} (\bibinfo {year} {2019})}\BibitemShut
  {NoStop}%
\bibitem [{\citenamefont {Rana}\ \emph {et~al.}(2023)\citenamefont {Rana},
  \citenamefont {Liao}, \citenamefont {Iacocca}, \citenamefont {Zou},
  \citenamefont {Pham}, \citenamefont {Lu}, \citenamefont {Subramanian},
  \citenamefont {Lo}, \citenamefont {Ryan}, \citenamefont {Bevis},
  \citenamefont {Karl}, \citenamefont {Glaid}, \citenamefont {Rable},
  \citenamefont {Mahale}, \citenamefont {Hirst}, \citenamefont {Ostler},
  \citenamefont {Liu}, \citenamefont {O’Leary}, \citenamefont {Yu},
  \citenamefont {Bustillo}, \citenamefont {Ohldag}, \citenamefont {Shapiro},
  \citenamefont {Yazdi}, \citenamefont {Mallouk}, \citenamefont {Osher},
  \citenamefont {Kapteyn}, \citenamefont {Crespi}, \citenamefont {Badding},
  \citenamefont {Tserkovnyak}, \citenamefont {Murnane},\ and\ \citenamefont
  {Miao}}]{2023Rana}%
  \BibitemOpen
  \bibfield  {author} {\bibinfo {author} {\bibfnamefont {A.}~\bibnamefont
  {Rana}}, \bibinfo {author} {\bibfnamefont {C.-T.}\ \bibnamefont {Liao}},
  \bibinfo {author} {\bibfnamefont {E.}~\bibnamefont {Iacocca}}, \bibinfo
  {author} {\bibfnamefont {J.}~\bibnamefont {Zou}}, \bibinfo {author}
  {\bibfnamefont {M.}~\bibnamefont {Pham}}, \bibinfo {author} {\bibfnamefont
  {X.}~\bibnamefont {Lu}}, \bibinfo {author} {\bibfnamefont {E.-E.~C.}\
  \bibnamefont {Subramanian}}, \bibinfo {author} {\bibfnamefont {Y.~H.}\
  \bibnamefont {Lo}}, \bibinfo {author} {\bibfnamefont {S.~A.}\ \bibnamefont
  {Ryan}}, \bibinfo {author} {\bibfnamefont {C.~S.}\ \bibnamefont {Bevis}},
  \bibinfo {author} {\bibfnamefont {R.~M.}\ \bibnamefont {Karl}}, \bibinfo
  {author} {\bibfnamefont {A.~J.}\ \bibnamefont {Glaid}}, \bibinfo {author}
  {\bibfnamefont {J.}~\bibnamefont {Rable}}, \bibinfo {author} {\bibfnamefont
  {P.}~\bibnamefont {Mahale}}, \bibinfo {author} {\bibfnamefont
  {J.}~\bibnamefont {Hirst}}, \bibinfo {author} {\bibfnamefont
  {T.}~\bibnamefont {Ostler}}, \bibinfo {author} {\bibfnamefont
  {W.}~\bibnamefont {Liu}}, \bibinfo {author} {\bibfnamefont {C.~M.}\
  \bibnamefont {O’Leary}}, \bibinfo {author} {\bibfnamefont {Y.-S.}\
  \bibnamefont {Yu}}, \bibinfo {author} {\bibfnamefont {K.}~\bibnamefont
  {Bustillo}}, \bibinfo {author} {\bibfnamefont {H.}~\bibnamefont {Ohldag}},
  \bibinfo {author} {\bibfnamefont {D.~A.}\ \bibnamefont {Shapiro}}, \bibinfo
  {author} {\bibfnamefont {S.}~\bibnamefont {Yazdi}}, \bibinfo {author}
  {\bibfnamefont {T.~E.}\ \bibnamefont {Mallouk}}, \bibinfo {author}
  {\bibfnamefont {S.~J.}\ \bibnamefont {Osher}}, \bibinfo {author}
  {\bibfnamefont {H.~C.}\ \bibnamefont {Kapteyn}}, \bibinfo {author}
  {\bibfnamefont {V.~H.}\ \bibnamefont {Crespi}}, \bibinfo {author}
  {\bibfnamefont {J.~V.}\ \bibnamefont {Badding}}, \bibinfo {author}
  {\bibfnamefont {Y.}~\bibnamefont {Tserkovnyak}}, \bibinfo {author}
  {\bibfnamefont {M.~M.}\ \bibnamefont {Murnane}},\ and\ \bibinfo {author}
  {\bibfnamefont {J.}~\bibnamefont {Miao}},\ }\bibfield  {title} {\bibinfo
  {title} {Three-dimensional topological magnetic monopoles and their
  interactions in a ferromagnetic meta-lattice},\ }\href
  {https://doi.org/10.1038/s41565-022-01311-0} {\bibfield  {journal} {\bibinfo
  {journal} {Nature Nanotechnology}\ }\textbf {\bibinfo {volume} {18}},\
  \bibinfo {pages} {227–232} (\bibinfo {year} {2023})}\BibitemShut {NoStop}%
\bibitem [{\citenamefont {Cheenikundil}\ and\ \citenamefont
  {Hertel}(2021)}]{2021Cheenikundil}%
  \BibitemOpen
  \bibfield  {author} {\bibinfo {author} {\bibfnamefont {R.}~\bibnamefont
  {Cheenikundil}}\ and\ \bibinfo {author} {\bibfnamefont {R.}~\bibnamefont
  {Hertel}},\ }\bibfield  {title} {\bibinfo {title} {{Switchable magnetic
  frustration in buckyball nanoarchitectures}},\ }\href
  {https://doi.org/10.1063/5.0048936} {\bibfield  {journal} {\bibinfo
  {journal} {Applied Physics Letters}\ }\textbf {\bibinfo {volume} {118}},\
  \bibinfo {pages} {212403} (\bibinfo {year} {2021})}\BibitemShut {NoStop}%
\bibitem [{\citenamefont {Turner}\ \emph {et~al.}(2013)\citenamefont {Turner},
  \citenamefont {Saba}, \citenamefont {Zhang}, \citenamefont {Cumming},
  \citenamefont {Schr{\"{o}}der-Turk},\ and\ \citenamefont {Gu}}]{2013Turner}%
  \BibitemOpen
  \bibfield  {author} {\bibinfo {author} {\bibfnamefont {M.~D.}\ \bibnamefont
  {Turner}}, \bibinfo {author} {\bibfnamefont {M.}~\bibnamefont {Saba}},
  \bibinfo {author} {\bibfnamefont {Q.}~\bibnamefont {Zhang}}, \bibinfo
  {author} {\bibfnamefont {B.~P.}\ \bibnamefont {Cumming}}, \bibinfo {author}
  {\bibfnamefont {G.~E.}\ \bibnamefont {Schr{\"{o}}der-Turk}},\ and\ \bibinfo
  {author} {\bibfnamefont {M.}~\bibnamefont {Gu}},\ }\bibfield  {title}
  {\bibinfo {title} {{Miniature chiral beamsplitter based on gyroid photonic
  crystals}},\ }\href {https://doi.org/10.1038/nphoton.2013.233} {\bibfield
  {journal} {\bibinfo  {journal} {Nature Photonics}\ }\textbf {\bibinfo
  {volume} {7}},\ \bibinfo {pages} {801} (\bibinfo {year} {2013})}\BibitemShut
  {NoStop}%
\bibitem [{\citenamefont {Koshino}\ and\ \citenamefont
  {Aoki}(2005)}]{2005Koshino}%
  \BibitemOpen
  \bibfield  {author} {\bibinfo {author} {\bibfnamefont {M.}~\bibnamefont
  {Koshino}}\ and\ \bibinfo {author} {\bibfnamefont {H.}~\bibnamefont {Aoki}},\
  }\bibfield  {title} {\bibinfo {title} {{Electronic structure of an electron
  on the gyroid surface: A helical labyrinth}},\ }\href
  {https://doi.org/10.1103/PhysRevB.71.073405} {\bibfield  {journal} {\bibinfo
  {journal} {Physical Review B}\ }\textbf {\bibinfo {volume} {71}},\ \bibinfo
  {pages} {073405} (\bibinfo {year} {2005})}\BibitemShut {NoStop}%
\bibitem [{\citenamefont {Lu}\ \emph {et~al.}(2013)\citenamefont {Lu},
  \citenamefont {Fu}, \citenamefont {Joannopoulos},\ and\ \citenamefont
  {Soljacic}}]{2013Lu}%
  \BibitemOpen
  \bibfield  {author} {\bibinfo {author} {\bibfnamefont {L.}~\bibnamefont
  {Lu}}, \bibinfo {author} {\bibfnamefont {L.}~\bibnamefont {Fu}}, \bibinfo
  {author} {\bibfnamefont {J.~D.}\ \bibnamefont {Joannopoulos}},\ and\ \bibinfo
  {author} {\bibfnamefont {M.}~\bibnamefont {Soljacic}},\ }\bibfield  {title}
  {\bibinfo {title} {Weyl points and line nodes in gyroid photonic crystals},\
  }\href {https://doi.org/10.1038/nphoton.2013.42} {\bibfield  {journal}
  {\bibinfo  {journal} {Nat Photon}\ }\textbf {\bibinfo {volume} {7}},\
  \bibinfo {pages} {294} (\bibinfo {year} {2013})}\BibitemShut {NoStop}%
\bibitem [{\citenamefont {Lu}\ \emph {et~al.}(2015)\citenamefont {Lu},
  \citenamefont {Wang}, \citenamefont {Ye}, \citenamefont {Ran}, \citenamefont
  {Fu}, \citenamefont {Joannopoulos},\ and\ \citenamefont
  {Soljačić}}]{2015Lu}%
  \BibitemOpen
  \bibfield  {author} {\bibinfo {author} {\bibfnamefont {L.}~\bibnamefont
  {Lu}}, \bibinfo {author} {\bibfnamefont {Z.}~\bibnamefont {Wang}}, \bibinfo
  {author} {\bibfnamefont {D.}~\bibnamefont {Ye}}, \bibinfo {author}
  {\bibfnamefont {L.}~\bibnamefont {Ran}}, \bibinfo {author} {\bibfnamefont
  {L.}~\bibnamefont {Fu}}, \bibinfo {author} {\bibfnamefont {J.~D.}\
  \bibnamefont {Joannopoulos}},\ and\ \bibinfo {author} {\bibfnamefont
  {M.}~\bibnamefont {Soljačić}},\ }\bibfield  {title} {\bibinfo {title}
  {Experimental observation of {Weyl} points},\ }\href
  {https://doi.org/10.1126/science.aaa9273} {\bibfield  {journal} {\bibinfo
  {journal} {Science}\ }\textbf {\bibinfo {volume} {349}},\ \bibinfo {pages}
  {622} (\bibinfo {year} {2015})}\BibitemShut {NoStop}%
\bibitem [{\citenamefont {Seki}\ \emph {et~al.}(2016)\citenamefont {Seki},
  \citenamefont {Okamura}, \citenamefont {Kondou}, \citenamefont {Shibata},
  \citenamefont {Kubota}, \citenamefont {Takagi}, \citenamefont {Kagawa},
  \citenamefont {Kawasaki}, \citenamefont {Tatara}, \citenamefont {Otani},\
  and\ \citenamefont {Tokura}}]{2016Seki}%
  \BibitemOpen
  \bibfield  {author} {\bibinfo {author} {\bibfnamefont {S.}~\bibnamefont
  {Seki}}, \bibinfo {author} {\bibfnamefont {Y.}~\bibnamefont {Okamura}},
  \bibinfo {author} {\bibfnamefont {K.}~\bibnamefont {Kondou}}, \bibinfo
  {author} {\bibfnamefont {K.}~\bibnamefont {Shibata}}, \bibinfo {author}
  {\bibfnamefont {M.}~\bibnamefont {Kubota}}, \bibinfo {author} {\bibfnamefont
  {R.}~\bibnamefont {Takagi}}, \bibinfo {author} {\bibfnamefont
  {F.}~\bibnamefont {Kagawa}}, \bibinfo {author} {\bibfnamefont
  {M.}~\bibnamefont {Kawasaki}}, \bibinfo {author} {\bibfnamefont
  {G.}~\bibnamefont {Tatara}}, \bibinfo {author} {\bibfnamefont
  {Y.}~\bibnamefont {Otani}},\ and\ \bibinfo {author} {\bibfnamefont
  {Y.}~\bibnamefont {Tokura}},\ }\bibfield  {title} {\bibinfo {title}
  {Magnetochiral nonreciprocity of volume spin wave propagation in
  chiral-lattice ferromagnets},\ }\href
  {https://doi.org/10.1103/PhysRevB.93.235131} {\bibfield  {journal} {\bibinfo
  {journal} {Phys. Rev. B}\ }\textbf {\bibinfo {volume} {93}},\ \bibinfo
  {pages} {235131} (\bibinfo {year} {2016})}\BibitemShut {NoStop}%
\bibitem [{\citenamefont {Tokura}\ and\ \citenamefont
  {Nagaosa}(2018)}]{2018Tokura}%
  \BibitemOpen
  \bibfield  {author} {\bibinfo {author} {\bibfnamefont {Y.}~\bibnamefont
  {Tokura}}\ and\ \bibinfo {author} {\bibfnamefont {N.}~\bibnamefont
  {Nagaosa}},\ }\bibfield  {title} {\bibinfo {title} {Nonreciprocal responses
  from non-centrosymmetric quantum materials},\ }\href
  {https://doi.org/10.1038/s41467-018-05759-4} {\bibfield  {journal} {\bibinfo
  {journal} {Nature Communications}\ }\textbf {\bibinfo {volume} {9}},\
  \bibinfo {pages} {3740} (\bibinfo {year} {2018})}\BibitemShut {NoStop}%
\bibitem [{\citenamefont {Barman}\ \emph {et~al.}(2020)\citenamefont {Barman},
  \citenamefont {Mondal}, \citenamefont {Sahoo},\ and\ \citenamefont
  {De}}]{2020Barman}%
  \BibitemOpen
  \bibfield  {author} {\bibinfo {author} {\bibfnamefont {A.}~\bibnamefont
  {Barman}}, \bibinfo {author} {\bibfnamefont {S.}~\bibnamefont {Mondal}},
  \bibinfo {author} {\bibfnamefont {S.}~\bibnamefont {Sahoo}},\ and\ \bibinfo
  {author} {\bibfnamefont {A.}~\bibnamefont {De}},\ }\bibfield  {title}
  {\bibinfo {title} {Magnetization dynamics of nanoscale magnetic materials:
  {A} perspective},\ }\href {https://doi.org/10.1063/5.0023993} {\bibfield
  {journal} {\bibinfo  {journal} {Journal of Applied Physics}\ }\textbf
  {\bibinfo {volume} {128}},\ \bibinfo {pages} {170901} (\bibinfo {year}
  {2020})}\BibitemShut {NoStop}%
\bibitem [{\citenamefont {Rikken}\ \emph {et~al.}(2001)\citenamefont {Rikken},
  \citenamefont {F{\"{o}}lling},\ and\ \citenamefont {Wyder}}]{2001Rikken}%
  \BibitemOpen
  \bibfield  {author} {\bibinfo {author} {\bibfnamefont {G.}~\bibnamefont
  {Rikken}}, \bibinfo {author} {\bibfnamefont {J.}~\bibnamefont
  {F{\"{o}}lling}},\ and\ \bibinfo {author} {\bibfnamefont {P.}~\bibnamefont
  {Wyder}},\ }\bibfield  {title} {\bibinfo {title} {Electrical magnetochiral
  anisotropy},\ }\href {https://doi.org/10.1103/PhysRevLett.87.236602}
  {\bibfield  {journal} {\bibinfo  {journal} {Physical Review Letters}\
  }\textbf {\bibinfo {volume} {87}},\ \bibinfo {pages} {236602} (\bibinfo
  {year} {2001})}\BibitemShut {NoStop}%
\bibitem [{\citenamefont {Atzori}\ \emph {et~al.}(2021)\citenamefont {Atzori},
  \citenamefont {Train}, \citenamefont {Hillard}, \citenamefont {Avarvari},\
  and\ \citenamefont {Rikken}}]{2021Atzori}%
  \BibitemOpen
  \bibfield  {author} {\bibinfo {author} {\bibfnamefont {M.}~\bibnamefont
  {Atzori}}, \bibinfo {author} {\bibfnamefont {C.}~\bibnamefont {Train}},
  \bibinfo {author} {\bibfnamefont {E.~A.}\ \bibnamefont {Hillard}}, \bibinfo
  {author} {\bibfnamefont {N.}~\bibnamefont {Avarvari}},\ and\ \bibinfo
  {author} {\bibfnamefont {G.~L. J.~A.}\ \bibnamefont {Rikken}},\ }\bibfield
  {title} {\bibinfo {title} {Magneto-chiral anisotropy: {F}rom fundamentals to
  perspectives},\ }\href {https://doi.org/10.1002/chir.23361} {\bibfield
  {journal} {\bibinfo  {journal} {Chirality}\ }\textbf {\bibinfo {volume}
  {33}},\ \bibinfo {pages} {844} (\bibinfo {year} {2021})}\BibitemShut
  {NoStop}%
\bibitem [{\citenamefont {Llandro}\ \emph {et~al.}(2020)\citenamefont
  {Llandro}, \citenamefont {Love}, \citenamefont {Kovács}, \citenamefont
  {Caron}, \citenamefont {Vyas}, \citenamefont {Kákay}, \citenamefont
  {Salikhov}, \citenamefont {Lenz}, \citenamefont {Fassbender}, \citenamefont
  {Scherer}, \citenamefont {Cimorra}, \citenamefont {Steiner}, \citenamefont
  {Barnes}, \citenamefont {Dunin-Borkowski}, \citenamefont {Fukami},\ and\
  \citenamefont {Ohno}}]{2020Llandro}%
  \BibitemOpen
  \bibfield  {author} {\bibinfo {author} {\bibfnamefont {J.}~\bibnamefont
  {Llandro}}, \bibinfo {author} {\bibfnamefont {D.~M.}\ \bibnamefont {Love}},
  \bibinfo {author} {\bibfnamefont {A.}~\bibnamefont {Kovács}}, \bibinfo
  {author} {\bibfnamefont {J.}~\bibnamefont {Caron}}, \bibinfo {author}
  {\bibfnamefont {K.~N.}\ \bibnamefont {Vyas}}, \bibinfo {author}
  {\bibfnamefont {A.}~\bibnamefont {Kákay}}, \bibinfo {author} {\bibfnamefont
  {R.}~\bibnamefont {Salikhov}}, \bibinfo {author} {\bibfnamefont
  {K.}~\bibnamefont {Lenz}}, \bibinfo {author} {\bibfnamefont {J.}~\bibnamefont
  {Fassbender}}, \bibinfo {author} {\bibfnamefont {M.~R.~J.}\ \bibnamefont
  {Scherer}}, \bibinfo {author} {\bibfnamefont {C.}~\bibnamefont {Cimorra}},
  \bibinfo {author} {\bibfnamefont {U.}~\bibnamefont {Steiner}}, \bibinfo
  {author} {\bibfnamefont {C.~H.~W.}\ \bibnamefont {Barnes}}, \bibinfo {author}
  {\bibfnamefont {R.~E.}\ \bibnamefont {Dunin-Borkowski}}, \bibinfo {author}
  {\bibfnamefont {S.}~\bibnamefont {Fukami}},\ and\ \bibinfo {author}
  {\bibfnamefont {H.}~\bibnamefont {Ohno}},\ }\bibfield  {title} {\bibinfo
  {title} {Visualizing magnetic structure in {3D} nanoscale {Ni-Fe} gyroid
  networks},\ }\href {https://doi.org/10.1021/acs.nanolett.0c00578} {\bibfield
  {journal} {\bibinfo  {journal} {Nano Lett.}\ }\textbf {\bibinfo {volume}
  {20}},\ \bibinfo {pages} {3642} (\bibinfo {year} {2020})}\BibitemShut
  {NoStop}%
\bibitem [{\citenamefont {Schoen}(1970)}]{1970Schoen}%
  \BibitemOpen
  \bibfield  {author} {\bibinfo {author} {\bibfnamefont {A.~H.}\ \bibnamefont
  {Schoen}},\ }\href@noop {} {\emph {\bibinfo {title} {Infinite periodic
  minimal surfaces without self-intersections}}},\ \bibinfo {type} {Tech.
  Rep.}\ (\bibinfo  {institution} {{NASA Electronics Research Center Cambridge,
  MA, United States}},\ \bibinfo {year} {1970})\BibitemShut {NoStop}%
\bibitem [{\citenamefont {Michielsen}\ and\ \citenamefont
  {Stavenga}(2007)}]{2007Michielsen}%
  \BibitemOpen
  \bibfield  {author} {\bibinfo {author} {\bibfnamefont {K.}~\bibnamefont
  {Michielsen}}\ and\ \bibinfo {author} {\bibfnamefont {D.~G.}\ \bibnamefont
  {Stavenga}},\ }\bibfield  {title} {\bibinfo {title} {Gyroid cuticular
  structures in butterfly wing scales: biological photonic crystals},\ }\href
  {https://doi.org/10.1098/rsif.2007.1065} {\bibfield  {journal} {\bibinfo
  {journal} {Journal of The Royal Society Interface}\ }\textbf {\bibinfo
  {volume} {5}},\ \bibinfo {pages} {85} (\bibinfo {year} {2007})}\BibitemShut
  {NoStop}%
\bibitem [{\citenamefont {Saranathan}\ \emph {et~al.}(2010)\citenamefont
  {Saranathan}, \citenamefont {Osuji}, \citenamefont {Mochrie}, \citenamefont
  {Noh}, \citenamefont {Narayanan}, \citenamefont {Sandy}, \citenamefont
  {Dufresne},\ and\ \citenamefont {Prum}}]{2010Saranathan}%
  \BibitemOpen
  \bibfield  {author} {\bibinfo {author} {\bibfnamefont {V.}~\bibnamefont
  {Saranathan}}, \bibinfo {author} {\bibfnamefont {C.~O.}\ \bibnamefont
  {Osuji}}, \bibinfo {author} {\bibfnamefont {S.~G.~J.}\ \bibnamefont
  {Mochrie}}, \bibinfo {author} {\bibfnamefont {H.}~\bibnamefont {Noh}},
  \bibinfo {author} {\bibfnamefont {S.}~\bibnamefont {Narayanan}}, \bibinfo
  {author} {\bibfnamefont {A.}~\bibnamefont {Sandy}}, \bibinfo {author}
  {\bibfnamefont {E.~R.}\ \bibnamefont {Dufresne}},\ and\ \bibinfo {author}
  {\bibfnamefont {R.~O.}\ \bibnamefont {Prum}},\ }\bibfield  {title} {\bibinfo
  {title} {Structure, function, and self-assembly of single network gyroid
  {$I4_132$} photonic crystals in butterfly wing scales},\ }\href
  {https://doi.org/10.1073/pnas.0909616107} {\bibfield  {journal} {\bibinfo
  {journal} {Proceedings of the National Academy of Sciences}\ }\textbf
  {\bibinfo {volume} {107}},\ \bibinfo {pages} {11676} (\bibinfo {year}
  {2010})}\BibitemShut {NoStop}%
\bibitem [{\citenamefont {Vignolini}\ \emph {et~al.}(2012)\citenamefont
  {Vignolini}, \citenamefont {Yufa}, \citenamefont {Cunha}, \citenamefont
  {Guldin}, \citenamefont {Rushkin}, \citenamefont {Stefik}, \citenamefont
  {Hur}, \citenamefont {Wiesner}, \citenamefont {Baumberg},\ and\ \citenamefont
  {Steiner}}]{2012Vignolini}%
  \BibitemOpen
  \bibfield  {author} {\bibinfo {author} {\bibfnamefont {S.}~\bibnamefont
  {Vignolini}}, \bibinfo {author} {\bibfnamefont {N.~A.}\ \bibnamefont {Yufa}},
  \bibinfo {author} {\bibfnamefont {P.~S.}\ \bibnamefont {Cunha}}, \bibinfo
  {author} {\bibfnamefont {S.}~\bibnamefont {Guldin}}, \bibinfo {author}
  {\bibfnamefont {I.}~\bibnamefont {Rushkin}}, \bibinfo {author} {\bibfnamefont
  {M.}~\bibnamefont {Stefik}}, \bibinfo {author} {\bibfnamefont
  {K.}~\bibnamefont {Hur}}, \bibinfo {author} {\bibfnamefont {U.}~\bibnamefont
  {Wiesner}}, \bibinfo {author} {\bibfnamefont {J.~J.}\ \bibnamefont
  {Baumberg}},\ and\ \bibinfo {author} {\bibfnamefont {U.}~\bibnamefont
  {Steiner}},\ }\bibfield  {title} {\bibinfo {title} {A 3d optical metamaterial
  made by self-assembly},\ }\href {https://doi.org/10.1002/adma.201103610}
  {\bibfield  {journal} {\bibinfo  {journal} {Adv. Mater.}\ }\textbf {\bibinfo
  {volume} {24}},\ \bibinfo {pages} {OP23} (\bibinfo {year}
  {2012})}\BibitemShut {NoStop}%
\bibitem [{\citenamefont {Dolan}\ \emph {et~al.}(2015)\citenamefont {Dolan},
  \citenamefont {Wilts}, \citenamefont {Vignolini}, \citenamefont {Baumberg},
  \citenamefont {Steiner},\ and\ \citenamefont {Wilkinson}}]{2015Dolan}%
  \BibitemOpen
  \bibfield  {author} {\bibinfo {author} {\bibfnamefont {J.~A.}\ \bibnamefont
  {Dolan}}, \bibinfo {author} {\bibfnamefont {B.~D.}\ \bibnamefont {Wilts}},
  \bibinfo {author} {\bibfnamefont {S.}~\bibnamefont {Vignolini}}, \bibinfo
  {author} {\bibfnamefont {J.~J.}\ \bibnamefont {Baumberg}}, \bibinfo {author}
  {\bibfnamefont {U.}~\bibnamefont {Steiner}},\ and\ \bibinfo {author}
  {\bibfnamefont {T.~D.}\ \bibnamefont {Wilkinson}},\ }\bibfield  {title}
  {\bibinfo {title} {Optical properties of gyroid structured materials: {F}rom
  photonic crystals to metamaterials},\ }\href
  {https://doi.org/10.1002/adom.201400333} {\bibfield  {journal} {\bibinfo
  {journal} {Advanced Optical Materials}\ }\textbf {\bibinfo {volume} {3}},\
  \bibinfo {pages} {12} (\bibinfo {year} {2015})}\BibitemShut {NoStop}%
\bibitem [{\citenamefont {Wu}\ \emph {et~al.}(2015)\citenamefont {Wu},
  \citenamefont {Zhang},\ and\ \citenamefont {Zhang}}]{2015Wu_a}%
  \BibitemOpen
  \bibfield  {author} {\bibinfo {author} {\bibfnamefont {L.}~\bibnamefont
  {Wu}}, \bibinfo {author} {\bibfnamefont {W.}~\bibnamefont {Zhang}},\ and\
  \bibinfo {author} {\bibfnamefont {D.}~\bibnamefont {Zhang}},\ }\bibfield
  {title} {\bibinfo {title} {Engineering gyroid-structured functional materials
  via templates discovered in nature and in the lab},\ }\href
  {https://doi.org/10.1002/smll.201500812} {\bibfield  {journal} {\bibinfo
  {journal} {Small}\ }\textbf {\bibinfo {volume} {11}},\ \bibinfo {pages}
  {5004} (\bibinfo {year} {2015})}\BibitemShut {NoStop}%
\bibitem [{\citenamefont {Dolan}\ \emph {et~al.}(2018)\citenamefont {Dolan},
  \citenamefont {Korzeb}, \citenamefont {Dehmel}, \citenamefont {Gödel},
  \citenamefont {Stefik}, \citenamefont {Wiesner}, \citenamefont {Wilkinson},
  \citenamefont {Baumberg}, \citenamefont {Wilts}, \citenamefont {Steiner},\
  and\ \citenamefont {Gunkel}}]{2018Dolan}%
  \BibitemOpen
  \bibfield  {author} {\bibinfo {author} {\bibfnamefont {J.~A.}\ \bibnamefont
  {Dolan}}, \bibinfo {author} {\bibfnamefont {K.}~\bibnamefont {Korzeb}},
  \bibinfo {author} {\bibfnamefont {R.}~\bibnamefont {Dehmel}}, \bibinfo
  {author} {\bibfnamefont {K.~C.}\ \bibnamefont {Gödel}}, \bibinfo {author}
  {\bibfnamefont {M.}~\bibnamefont {Stefik}}, \bibinfo {author} {\bibfnamefont
  {U.}~\bibnamefont {Wiesner}}, \bibinfo {author} {\bibfnamefont {T.~D.}\
  \bibnamefont {Wilkinson}}, \bibinfo {author} {\bibfnamefont {J.~J.}\
  \bibnamefont {Baumberg}}, \bibinfo {author} {\bibfnamefont {B.~D.}\
  \bibnamefont {Wilts}}, \bibinfo {author} {\bibfnamefont {U.}~\bibnamefont
  {Steiner}},\ and\ \bibinfo {author} {\bibfnamefont {I.}~\bibnamefont
  {Gunkel}},\ }\bibfield  {title} {\bibinfo {title} {Controlling self-assembly
  in gyroid terpolymer films by solvent vapor annealing},\ }\href
  {https://doi.org/https://doi.org/10.1002/smll.201802401} {\bibfield
  {journal} {\bibinfo  {journal} {Small}\ }\textbf {\bibinfo {volume} {14}},\
  \bibinfo {pages} {1802401} (\bibinfo {year} {2018})}\BibitemShut {NoStop}%
\bibitem [{\citenamefont {Hyde}\ \emph {et~al.}(2008)\citenamefont {Hyde},
  \citenamefont {O'Keeffe},\ and\ \citenamefont
  {Proserpio}}]{2008Hyde_english}%
  \BibitemOpen
  \bibfield  {author} {\bibinfo {author} {\bibfnamefont {S.~T.}\ \bibnamefont
  {Hyde}}, \bibinfo {author} {\bibfnamefont {M.}~\bibnamefont {O'Keeffe}},\
  and\ \bibinfo {author} {\bibfnamefont {D.~M.}\ \bibnamefont {Proserpio}},\
  }\bibfield  {title} {\bibinfo {title} {A short history of an elusive yet
  ubiquitous structure in chemistry, materials, and mathematics},\ }\href
  {https://doi.org/10.1002/anie.200801519} {\bibfield  {journal} {\bibinfo
  {journal} {Angewandte Chemie International Edition}\ }\textbf {\bibinfo
  {volume} {47}},\ \bibinfo {pages} {7996} (\bibinfo {year}
  {2008})}\BibitemShut {NoStop}%
\bibitem [{\citenamefont {Mizuno}\ \emph {et~al.}(2019)\citenamefont {Mizuno},
  \citenamefont {Shuku},\ and\ \citenamefont {Awaga}}]{2019Mizuno}%
  \BibitemOpen
  \bibfield  {author} {\bibinfo {author} {\bibfnamefont {A.}~\bibnamefont
  {Mizuno}}, \bibinfo {author} {\bibfnamefont {Y.}~\bibnamefont {Shuku}},\ and\
  \bibinfo {author} {\bibfnamefont {K.}~\bibnamefont {Awaga}},\ }\bibfield
  {title} {\bibinfo {title} {Recent developments in molecular spin gyroid
  research},\ }\href {https://doi.org/10.1246/bcsj.20190033} {\bibfield
  {journal} {\bibinfo  {journal} {Bulletin of the Chemical Society of Japan}\
  }\textbf {\bibinfo {volume} {92}},\ \bibinfo {pages} {1068} (\bibinfo {year}
  {2019})}\BibitemShut {NoStop}%
\bibitem [{\citenamefont {Wannier}(1950)}]{1950Wannier}%
  \BibitemOpen
  \bibfield  {author} {\bibinfo {author} {\bibfnamefont {G.~H.}\ \bibnamefont
  {Wannier}},\ }\bibfield  {title} {\bibinfo {title} {Antiferromagnetism. {The}
  triangular {Ising} net},\ }\href {https://doi.org/10.1103/PhysRev.79.357}
  {\bibfield  {journal} {\bibinfo  {journal} {Phys. Rev.}\ }\textbf {\bibinfo
  {volume} {79}},\ \bibinfo {pages} {357} (\bibinfo {year} {1950})}\BibitemShut
  {NoStop}%
\bibitem [{\citenamefont {Diep}(2004)}]{2004Diep}%
  \BibitemOpen
  \bibinfo {editor} {\bibfnamefont {H.}~\bibnamefont {Diep}},\ ed.,\ \href@noop
  {} {\emph {\bibinfo {title} {Frustrated Spin Systems}}}\ (\bibinfo
  {publisher} {World Scientific},\ \bibinfo {year} {2004})\BibitemShut
  {NoStop}%
\bibitem [{\citenamefont {Lacroix}\ \emph {et~al.}(2011)\citenamefont
  {Lacroix}, \citenamefont {Mendels},\ and\ \citenamefont {Mila}}]{Lacroix11}%
  \BibitemOpen
  \bibfield  {author} {\bibinfo {author} {\bibfnamefont {C.}~\bibnamefont
  {Lacroix}}, \bibinfo {author} {\bibfnamefont {P.}~\bibnamefont {Mendels}},\
  and\ \bibinfo {author} {\bibfnamefont {F.}~\bibnamefont {Mila}},\ }\href@noop
  {} {\emph {\bibinfo {title} {Introduction to Frustrated Magnetism}}}\
  (\bibinfo  {publisher} {Springer},\ \bibinfo {year} {2011})\BibitemShut
  {NoStop}%
\bibitem [{\citenamefont {Bisotti}\ \emph {et~al.}(2018)\citenamefont
  {Bisotti}, \citenamefont {Beg}, \citenamefont {Wang}, \citenamefont {Albert},
  \citenamefont {Chernyshenko}, \citenamefont {Cort{\'{e}}s-Ortu{\~{n}}o},
  \citenamefont {Pepper}, \citenamefont {Vousden}, \citenamefont {Carey},
  \citenamefont {Fuchs}, \citenamefont {Johansen}, \citenamefont {Balaban},
  \citenamefont {Breth}, \citenamefont {Kluyver},\ and\ \citenamefont
  {Fangohr}}]{2018finmag}%
  \BibitemOpen
  \bibfield  {author} {\bibinfo {author} {\bibfnamefont {M.}~\bibnamefont
  {Bisotti}}, \bibinfo {author} {\bibfnamefont {M.}~\bibnamefont {Beg}},
  \bibinfo {author} {\bibfnamefont {W.}~\bibnamefont {Wang}}, \bibinfo {author}
  {\bibfnamefont {M.}~\bibnamefont {Albert}}, \bibinfo {author} {\bibfnamefont
  {D.}~\bibnamefont {Chernyshenko}}, \bibinfo {author} {\bibfnamefont
  {D.}~\bibnamefont {Cort{\'{e}}s-Ortu{\~{n}}o}}, \bibinfo {author}
  {\bibfnamefont {R.~A.}\ \bibnamefont {Pepper}}, \bibinfo {author}
  {\bibfnamefont {M.}~\bibnamefont {Vousden}}, \bibinfo {author} {\bibfnamefont
  {R.}~\bibnamefont {Carey}}, \bibinfo {author} {\bibfnamefont
  {H.}~\bibnamefont {Fuchs}}, \bibinfo {author} {\bibfnamefont
  {A.}~\bibnamefont {Johansen}}, \bibinfo {author} {\bibfnamefont
  {G.}~\bibnamefont {Balaban}}, \bibinfo {author} {\bibfnamefont
  {L.}~\bibnamefont {Breth}}, \bibinfo {author} {\bibfnamefont
  {T.}~\bibnamefont {Kluyver}},\ and\ \bibinfo {author} {\bibfnamefont
  {H.}~\bibnamefont {Fangohr}},\ }\href
  {https://doi.org/10.5281/zenodo.1216011} {\bibinfo {title} {{FinMag:
  {Finite}-element micromagnetic simulation tool}}} (\bibinfo {year}
  {2018})\BibitemShut {NoStop}%
\bibitem [{\citenamefont {{COMSOL
  Multiphysics\textsuperscript{\textregistered}}}(2019)}]{comsol}%
  \BibitemOpen
  \bibfield  {author} {\bibinfo {author} {\bibnamefont {{COMSOL
  Multiphysics\textsuperscript{\textregistered}}}},\ }\href
  {http://www.comsol.com} {\emph {\bibinfo {title} {{v. 5.3}}}}\ (\bibinfo
  {publisher} {COMSOL AB},\ \bibinfo {address} {Stockholm, Sweden},\ \bibinfo
  {year} {2019})\BibitemShut {NoStop}%
\bibitem [{\citenamefont {Hagberg}\ \emph {et~al.}(2008)\citenamefont
  {Hagberg}, \citenamefont {Swart},\ and\ \citenamefont {S~Chult}}]{networkx}%
  \BibitemOpen
  \bibfield  {author} {\bibinfo {author} {\bibfnamefont {A.}~\bibnamefont
  {Hagberg}}, \bibinfo {author} {\bibfnamefont {P.}~\bibnamefont {Swart}},\
  and\ \bibinfo {author} {\bibfnamefont {D.}~\bibnamefont {S~Chult}},\
  }\bibfield  {title} {\bibinfo {title} {Exploring network structure, dynamics,
  and function using networkx}\ }(\bibinfo {year} {2008})\BibitemShut {NoStop}%
\bibitem [{\citenamefont {Taguchi}\ \emph {et~al.}(2001)\citenamefont
  {Taguchi}, \citenamefont {Oohara}, \citenamefont {Yoshizawa}, \citenamefont
  {Nagaosa},\ and\ \citenamefont {Tokura}}]{2001Taguchi}%
  \BibitemOpen
  \bibfield  {author} {\bibinfo {author} {\bibfnamefont {Y.}~\bibnamefont
  {Taguchi}}, \bibinfo {author} {\bibfnamefont {Y.}~\bibnamefont {Oohara}},
  \bibinfo {author} {\bibfnamefont {H.}~\bibnamefont {Yoshizawa}}, \bibinfo
  {author} {\bibfnamefont {N.}~\bibnamefont {Nagaosa}},\ and\ \bibinfo {author}
  {\bibfnamefont {Y.}~\bibnamefont {Tokura}},\ }\bibfield  {title} {\bibinfo
  {title} {Spin chirality, {Berry} phase, and anomalous {Hall} effect in a
  frustrated ferromagnet},\ }\href {https://doi.org/10.1126/science.1058161}
  {\bibfield  {journal} {\bibinfo  {journal} {Science}\ }\textbf {\bibinfo
  {volume} {291}},\ \bibinfo {pages} {2573} (\bibinfo {year}
  {2001})}\BibitemShut {NoStop}%
\bibitem [{\citenamefont {Park}\ \emph {et~al.}(2008)\citenamefont {Park},
  \citenamefont {Onoda}, \citenamefont {Nagaosa},\ and\ \citenamefont
  {Han}}]{2008Park}%
  \BibitemOpen
  \bibfield  {author} {\bibinfo {author} {\bibfnamefont {J.-H.}\ \bibnamefont
  {Park}}, \bibinfo {author} {\bibfnamefont {S.}~\bibnamefont {Onoda}},
  \bibinfo {author} {\bibfnamefont {N.}~\bibnamefont {Nagaosa}},\ and\ \bibinfo
  {author} {\bibfnamefont {J.~H.}\ \bibnamefont {Han}},\ }\bibfield  {title}
  {\bibinfo {title} {Nematic and chiral order for planar spins on a triangular
  lattice},\ }\href {https://doi.org/10.1103/PhysRevLett.101.167202} {\bibfield
   {journal} {\bibinfo  {journal} {Phys. Rev. Lett.}\ }\textbf {\bibinfo
  {volume} {101}},\ \bibinfo {pages} {167202} (\bibinfo {year}
  {2008})}\BibitemShut {NoStop}%
\bibitem [{\citenamefont {Wang}\ \emph {et~al.}(2005)\citenamefont {Wang},
  \citenamefont {Adeyeye}, \citenamefont {Singh}, \citenamefont {Huang},\ and\
  \citenamefont {Wu}}]{2005Wang}%
  \BibitemOpen
  \bibfield  {author} {\bibinfo {author} {\bibfnamefont {C.~C.}\ \bibnamefont
  {Wang}}, \bibinfo {author} {\bibfnamefont {A.~O.}\ \bibnamefont {Adeyeye}},
  \bibinfo {author} {\bibfnamefont {N.}~\bibnamefont {Singh}}, \bibinfo
  {author} {\bibfnamefont {Y.~S.}\ \bibnamefont {Huang}},\ and\ \bibinfo
  {author} {\bibfnamefont {Y.~H.}\ \bibnamefont {Wu}},\ }\bibfield  {title}
  {\bibinfo {title} {Magnetoresistance behavior of nanoscale antidot arrays},\
  }\href {https://doi.org/10.1103/PhysRevB.72.174426} {\bibfield  {journal}
  {\bibinfo  {journal} {Phys. Rev. B}\ }\textbf {\bibinfo {volume} {72}},\
  \bibinfo {pages} {174426} (\bibinfo {year} {2005})}\BibitemShut {NoStop}%
\bibitem [{\citenamefont {Pignard}\ \emph {et~al.}(1999)\citenamefont
  {Pignard}, \citenamefont {Goglio}, \citenamefont {Radulescu}, \citenamefont
  {Piraux}, \citenamefont {Dubois}, \citenamefont {Declémy},\ and\
  \citenamefont {Duvail}}]{2000Pignard}%
  \BibitemOpen
  \bibfield  {author} {\bibinfo {author} {\bibfnamefont {S.}~\bibnamefont
  {Pignard}}, \bibinfo {author} {\bibfnamefont {G.}~\bibnamefont {Goglio}},
  \bibinfo {author} {\bibfnamefont {A.}~\bibnamefont {Radulescu}}, \bibinfo
  {author} {\bibfnamefont {L.}~\bibnamefont {Piraux}}, \bibinfo {author}
  {\bibfnamefont {S.}~\bibnamefont {Dubois}}, \bibinfo {author} {\bibfnamefont
  {A.}~\bibnamefont {Declémy}},\ and\ \bibinfo {author} {\bibfnamefont
  {J.~L.}\ \bibnamefont {Duvail}},\ }\bibfield  {title} {\bibinfo {title}
  {Study of the magnetization reversal in individual nickel nanowires},\ }\href
  {https://doi.org/10.1063/1.371947} {\bibfield  {journal} {\bibinfo  {journal}
  {Journal of Applied Physics}\ }\textbf {\bibinfo {volume} {87}},\ \bibinfo
  {pages} {824} (\bibinfo {year} {1999})}\BibitemShut {NoStop}%
\bibitem [{\citenamefont {Klein}\ and\ \citenamefont
  {Randić}(1993)}]{1993Klein}%
  \BibitemOpen
  \bibfield  {author} {\bibinfo {author} {\bibfnamefont {D.~J.}\ \bibnamefont
  {Klein}}\ and\ \bibinfo {author} {\bibfnamefont {M.}~\bibnamefont
  {Randić}},\ }\bibfield  {title} {\bibinfo {title} {Resistance distance},\
  }\href {https://doi.org/10.1007/BF01164627} {\bibfield  {journal} {\bibinfo
  {journal} {Journal of Mathematical Chemistry}\ }\textbf {\bibinfo {volume}
  {12}},\ \bibinfo {pages} {81} (\bibinfo {year} {1993})}\BibitemShut {NoStop}%
\bibitem [{\citenamefont {Fernández-Pacheco}\ \emph
  {et~al.}(2020)\citenamefont {Fernández-Pacheco}, \citenamefont {Skoric},
  \citenamefont {De~Teresa}, \citenamefont {Pablo-Navarro}, \citenamefont
  {Huth},\ and\ \citenamefont {Dobrovolskiy}}]{2020FernandezPacheco}%
  \BibitemOpen
  \bibfield  {author} {\bibinfo {author} {\bibfnamefont {A.}~\bibnamefont
  {Fernández-Pacheco}}, \bibinfo {author} {\bibfnamefont {L.}~\bibnamefont
  {Skoric}}, \bibinfo {author} {\bibfnamefont {J.~M.}\ \bibnamefont
  {De~Teresa}}, \bibinfo {author} {\bibfnamefont {J.}~\bibnamefont
  {Pablo-Navarro}}, \bibinfo {author} {\bibfnamefont {M.}~\bibnamefont
  {Huth}},\ and\ \bibinfo {author} {\bibfnamefont {O.~V.}\ \bibnamefont
  {Dobrovolskiy}},\ }\href {https://doi.org/10.3390/ma13173774} {\bibinfo
  {title} {Writing {3D} nanomagnets using focused electron beams}} (\bibinfo
  {year} {2020})\BibitemShut {NoStop}%
\bibitem [{\citenamefont {Skoric}\ \emph {et~al.}(2020)\citenamefont {Skoric},
  \citenamefont {Sanz-Hernández}, \citenamefont {Meng}, \citenamefont
  {Donnelly}, \citenamefont {Merino-Aceituno},\ and\ \citenamefont
  {Fernández-Pacheco}}]{2020Skoric}%
  \BibitemOpen
  \bibfield  {author} {\bibinfo {author} {\bibfnamefont {L.}~\bibnamefont
  {Skoric}}, \bibinfo {author} {\bibfnamefont {D.}~\bibnamefont
  {Sanz-Hernández}}, \bibinfo {author} {\bibfnamefont {F.}~\bibnamefont
  {Meng}}, \bibinfo {author} {\bibfnamefont {C.}~\bibnamefont {Donnelly}},
  \bibinfo {author} {\bibfnamefont {S.}~\bibnamefont {Merino-Aceituno}},\ and\
  \bibinfo {author} {\bibfnamefont {A.}~\bibnamefont {Fernández-Pacheco}},\
  }\bibfield  {title} {\bibinfo {title} {Layer-by-layer growth of
  complex-shaped three-dimensional nanostructures with focused electron
  beams},\ }\href {https://doi.org/10.1021/acs.nanolett.9b03565} {\bibfield
  {journal} {\bibinfo  {journal} {Nano Lett.}\ }\textbf {\bibinfo {volume}
  {20}},\ \bibinfo {pages} {184} (\bibinfo {year} {2020})}\BibitemShut
  {NoStop}%
\bibitem [{\citenamefont {Williams}\ \emph {et~al.}(2018)\citenamefont
  {Williams}, \citenamefont {Hunt}, \citenamefont {Boehm}, \citenamefont {May},
  \citenamefont {Taverne}, \citenamefont {Ho}, \citenamefont {Giblin},
  \citenamefont {Read}, \citenamefont {Rarity}, \citenamefont {Allenspach},\
  and\ \citenamefont {Ladak}}]{2018Williams}%
  \BibitemOpen
  \bibfield  {author} {\bibinfo {author} {\bibfnamefont {G.}~\bibnamefont
  {Williams}}, \bibinfo {author} {\bibfnamefont {M.}~\bibnamefont {Hunt}},
  \bibinfo {author} {\bibfnamefont {B.}~\bibnamefont {Boehm}}, \bibinfo
  {author} {\bibfnamefont {A.}~\bibnamefont {May}}, \bibinfo {author}
  {\bibfnamefont {M.}~\bibnamefont {Taverne}}, \bibinfo {author} {\bibfnamefont
  {D.}~\bibnamefont {Ho}}, \bibinfo {author} {\bibfnamefont {S.}~\bibnamefont
  {Giblin}}, \bibinfo {author} {\bibfnamefont {D.}~\bibnamefont {Read}},
  \bibinfo {author} {\bibfnamefont {J.}~\bibnamefont {Rarity}}, \bibinfo
  {author} {\bibfnamefont {R.}~\bibnamefont {Allenspach}},\ and\ \bibinfo
  {author} {\bibfnamefont {S.}~\bibnamefont {Ladak}},\ }\bibfield  {title}
  {\bibinfo {title} {Two-photon lithography for {3D} magnetic nanostructure
  fabrication},\ }\href {https://doi.org/10.1007/s12274-017-1694-0} {\bibfield
  {journal} {\bibinfo  {journal} {Nano Research}\ }\textbf {\bibinfo {volume}
  {11}},\ \bibinfo {pages} {845} (\bibinfo {year} {2018})}\BibitemShut
  {NoStop}%
\bibitem [{\citenamefont {van~den Berg}\ \emph {et~al.}(2023)\citenamefont
  {van~den Berg}, \citenamefont {Caruel}, \citenamefont {Hunt},\ and\
  \citenamefont {Ladak}}]{2023Berg}%
  \BibitemOpen
  \bibfield  {author} {\bibinfo {author} {\bibfnamefont {A.}~\bibnamefont
  {van~den Berg}}, \bibinfo {author} {\bibfnamefont {M.}~\bibnamefont
  {Caruel}}, \bibinfo {author} {\bibfnamefont {M.}~\bibnamefont {Hunt}},\ and\
  \bibinfo {author} {\bibfnamefont {S.}~\bibnamefont {Ladak}},\ }\bibfield
  {title} {\bibinfo {title} {Combining two-photon lithography with laser
  ablation of sacrificial layers: {A} route to isolated {3D} magnetic
  nanostructures},\ }\href {https://doi.org/10.1007/s12274-022-4649-z}
  {\bibfield  {journal} {\bibinfo  {journal} {Nano Research}\ }\textbf
  {\bibinfo {volume} {16}},\ \bibinfo {pages} {1441} (\bibinfo {year}
  {2023})}\BibitemShut {NoStop}%
\bibitem [{\citenamefont {Fukami}\ and\ \citenamefont
  {Ohno}(2018)}]{2018Fukami}%
  \BibitemOpen
  \bibfield  {author} {\bibinfo {author} {\bibfnamefont {S.}~\bibnamefont
  {Fukami}}\ and\ \bibinfo {author} {\bibfnamefont {H.}~\bibnamefont {Ohno}},\
  }\bibfield  {title} {\bibinfo {title} {Perspective: Spintronic synapse for
  artificial neural network},\ }\href {https://doi.org/10.1063/1.5042317}
  {\bibfield  {journal} {\bibinfo  {journal} {Journal of Applied Physics}\
  }\textbf {\bibinfo {volume} {124}},\ \bibinfo {pages} {151904} (\bibinfo
  {year} {2018})} \BibitemShut {NoStop}%
	\bibitem [67]{data_repository}%
	\BibitemOpen
	\bibfield  {author} {A.S. Koshikawa, J. Llandro, M. Ohzeki, S. Fukami, H. Ohno, and N. Leo}\bibfield  {title} {\bibinfo {title} {data and code repository for "Magnetic Order in Nanoscale Gyroid Netwoks"},\ }\href {https://zenodo.org/records/10996003}
	{doi: 10.5281/zenodo.10996003 (\bibinfo
	{year} {2023})} \BibitemShut {NoStop}%
%
\end{thebibliography}

\end{document}